\documentclass[twoside,superscriptaddress,groupedaddress]{article}

\usepackage{ppnp}       
\usepackage{aas_macros} 
\usepackage{graphicx}   
\usepackage{bm}         
\usepackage{amssymb}    
\usepackage{amsmath}    
\usepackage{url}        
\usepackage{longtable } 
\usepackage{caption}    
\newcommand{\nggn}{$(n,\gamma)\rightleftarrows(\gamma,n)$}  

\begin{document}

\title{The impact of individual nuclear properties on $r$-process nucleosynthesis}

\author{M.~R. Mumpower,$^{1}$ R. Surman,$^{1}$ G.~C. McLaughlin,$^{2}$ A. Aprahamian,$^{1}$\\ 
\\
$^1$Department of Physics, University of Notre Dame, Notre Dame, IN 46556, USA\\
$^2$Department of Physics, North Carolina State University, Raleigh, NC 27695, USA\\
}

\maketitle

\begin{abstract}
The astrophysical rapid neutron capture process or `$r$ process' of nucleosynthesis is believed to be responsible for the production 
of approximately half the heavy element abundances found in nature. This multifaceted problem remains one of the greatest open 
challenges in all of physics. Knowledge of nuclear physics properties such as masses, $\beta$-decay and neutron capture rates, as well 
as $\beta$-delayed neutron emission probabilities are critical inputs that go into calculations of $r$-process nucleosynthesis. While 
properties of nuclei near stability have been established, much still remains unknown regarding neutron-rich nuclei far from 
stability that may participate in the $r$ process. Sensitivity studies gauge the astrophysical response of a change in nuclear 
physics input(s) which allows for the isolation of the most important nuclear properties that shape the final abundances observed in 
nature. This review summarizes the extent of recent sensitivity studies and highlights how these studies play a key role in 
facilitating new insight into the $r$ process. The development of these tools promotes a focused effort for state-of-the-art 
measurements, motivates construction of new facilities and will ultimately move the community towards addressing the grand challenge 
of `How were the elements from iron to uranium made?'.
\end{abstract}

\tableofcontents

\section{Introduction}
\label{sec:intro}

One of the major open questions in all of physics is the identification of the sites responsible for the production of the 
heaviest elements \cite{NRC03,NRC13}. It has been understood since the 1950’s that the solar system abundances of nuclei 
heavier than iron can be divided roughly in half based on the nucleosynthesis processes that create them. Slow neutron 
capture process, or $s$-process, nuclei lie along the middle of the valley of stability, and rapid neutron capture process, 
or $r$-process, nuclei are found on the neutron-rich side, with a third process, the $p$ process, responsible for the 
significantly less abundant nuclei on the proton-rich side of stability \cite{Burbidge+57,Cameron57}. Since that time much 
progress has been made, e.g. \cite{Wallerstein+97}, and the basic mechanisms of and astrophysical sites for the creation of 
the $s$-process \cite{Kappeler+11} and heavy $p$-process \cite{Arnould+03,Thielemann+10} nuclei are on a firm footing. The 
site or sites of the $r$ process still evade definitive determination \cite{Qian+07,Arnould+07,Thielemann+11}.

The $r$-process pattern is extracted from the solar system abundances by subtracting the $s$-process and 
$p$-process contributions \cite{Anders+88,Arlandini+99}. The residual pattern consists of three main abundance 
peaks at $A\sim 80$, $130$, and $195$, associated with the $N=50$, $82$, and $126$ closed shells. Building up to the 
heaviest $r$-process elements requires on the order of 100 neutrons per seed nucleus. Additional constraints come 
from meteoritic data, e.g. \cite{Wasserburg+96}, and observations of $r$-process elements in old stars in the 
galactic halo, e.g. \cite{Sneden+08,Roederer+14}. This data points to distinct origins for the light $A<120$ 
(`weak') and heavy $A>120$ (`main') $r$-process nuclei. The pattern of main $r$-process elements is remarkably 
similar among $r$-process enhanced halo stars and is a match to the solar residuals. This suggests that whatever 
mechanism created these elements must have been operating in a consistent fashion since early galactic times 
\cite{Mathews+90,Argast+04,Komiya+14,Matteucci+15}. Core-collapse supernovae fit the timescale argument, though 
the promise of early studies \cite{Meyer+92,Woosley+94} has not been achieved in modern simulations 
\cite{Arcones+07,Fischer+10,Hudepohl+10,Roberts+12}. Cold or mildly-heated outflows from neutron star mergers are 
more robustly neutron-rich \cite{Lattimer+74,Meyer89,Freiburghaus+99,Wanajo+14,Just+15}, potentially producing a 
consistent abundance pattern through fission recycling \cite{Goriely+11,Korobkin+12}. Mergers, however, suffer 
from an uncertain delay time \cite{Wanderman+15}. Other potential astrophysical sites that have been explored 
include hot accretion disk outflows from neutron star or neutron star-black hole mergers 
\cite{Surman+08,Perego+14,Wanajo+14,Just+15}, supernova neutron-rich jets \cite{Winteler+12}, gamma-ray burst 
collapsar outflows \cite{Pruet+03,Surman+06,Malkus+12}, shocked surface layers of O-Ne-Mg cores 
\cite{Wanajo+03,Janka+08}, and neutrino-induced nucleosynthesis in the helium shell of exploding massive stars 
\cite{Banerjee+11}.

The astrophysical sites listed above are all characterized by distinct conditions---temperature and density as a 
function of time, initial composition and neutron-richness---that should produce unique abundance pattern 
signatures. In principle, this should lead us directly to the astrophysical site of the main $r$ process, since 
this pattern is known to excellent precision and appears to be nearly universal \cite{Roederer+10}. However, the 
nuclear network calculations currently used to generate $r$-process predictions are still too uncertain for such 
detailed comparisons to be reliable. The uncertainties arise from the difficulties in modeling astrophysical 
environments as well as the unknown nuclear properties of the thousands of neutron-rich nuclear species that 
participate in the $r$ process.

Early studies of the $r$ process, reviewed in, e.g., \cite{Cowan+91}, noted the importance of nuclear masses and $\beta$-decay rates 
on $r$-process abundance predictions. These models assumed the $r$ process proceeds via rapid neutron captures on seed nuclei in a 
hot environment, such that neutron captures, $(n,\gamma)$, and photodissociations, $(\gamma,n)$, are in equilibrium. The abundances 
along an isotopic chain in \nggn \ equilibrium are set by the temperature, neutron abundance, and the neutron separation energies 
$S_{n}(Z,A)=E_{B}(Z,A)-E_{B}(Z,A-1)$ with the binding energy of a nucleus denoted by $E_{B}(Z,A)$. In this equilibrium picture, a 
steady $\beta$-flow is established between isotopic chains, with the relative abundances $Y$ set by the $\beta$-decay rates 
$\lambda_{\beta}$ of the most populated isotopes $(Z,A_{path})$, according to $\lambda_{\beta}(Z,A_{path})Y(Z,A_{path})\sim$ 
constant. The final abundance pattern is then obtained by considering $\beta$-delayed neutron emission during the decay back to 
stability, invoked to smooth out the strong odd-even effects in the equilibrium pattern \cite{Cameron+70,Kodama+75}. Modern 
$r$-process simulations use nuclear network codes with no assumption of \nggn \ equilibrium. These calculations confirm that for many 
astrophysical environments, \nggn \ equilibrium is indeed established at early times. At some point, however, this equilibrium fails, 
and the abundance pattern is finalized by an interplay between neutron captures, photodissociations, $\beta$-decays, and 
$\beta$-delayed neutron emission. In other astrophysical environments, equilibrium is established briefly if at all, and thus the 
full dynamics of the process rely on masses, reaction rates, $\beta$-decay lifetimes, and branching ratios for nuclei from the valley 
of stability to the neutron drip line. Beyond these basic inputs, more nuclear properties may be required depending on the 
astrophysical environment. The possibility of fission recycling includes estimates, e.g. of barrier heights and fragment 
distributions while in the case where neutrino fluxes are high, neutrino-nucleus reaction rates may be required.

Of the thousands of pieces of nuclear data required for main $r$-process simulations, to date few have been measured, given the 
experimental challenge of accessing nuclei far from stability. Simulations instead rely on theoretical models for the needed values. 
The vast majority of investigations into the role of nuclear data in the $r$ process have focused on the impacts of aspects of these 
models on $r$-process simulations. Arnould et al (2007) \cite{Arnould+07} provides a comprehensive review; additionally there are a 
wealth of recent studies of how models of nuclear mass and deformation 
\cite{Sun+08a,Sun+08b,Niu+09,Arcones+11,Arcones+12,Lahiri+12,Jiang+12,Soderstrom+12,Xu+13,Meng+13,Kratz+14}, $\beta$-decay lifetime 
predictions 
\cite{Vretenar+08,Famiano+08,Borzov+08,Marketin+09,Costiris+09,Borzov11,Suzuki+12,Niu+13,Fang+13,Lhersonneau+14,Caballero+14,Morales+14b,Severyukhin+14,Langanke+15,Suzuki15}, 
neutron capture rate calculations \cite{Chiba+08,Litvinova+09,Larsen+10,Zhang+12,Bertolli+14,Rauscher+15}, theoretical fission 
properties \cite{Panov+08,Minato+08,Korneev+11,Erler+12,Panov+13,Goriely+13,Eichler+14,Mendoza-Temis+14,Goriely15,Moller+15}, and neutrino-nucleus 
interaction rates \cite{Kajino13,Balasi+15} shape $r$-process abundance predictions.

The experimental situation, however, is evolving rapidly. Radioactive ion beam facilities are completely changing the landscape of 
accessible nuclei far from stability. Existing facilities and those coming on-line in the near future are expanding the scope of 
experimental nuclear science to the neutron rich regions of the chart of nuclides.  Recently nuclear masses of interest for the 
$r$-process have been measured with Penning traps at the ISOLDE facility at CERN \cite{Baruah+08}, JYFLTRAP at Jyv{\"a}skyl{\"a} 
\cite{Rahaman+07,Hakala+12}, and the CPT at CARIBU \cite{Savard+11,VanSchelt+13}; via time-of-flight (TOF) at the NSCL 
\cite{Matos+08}; and with the Fragment Separator (FRS) \cite{Benlliure+08} and via Isochronous Mass Spectrometry (IMS) at GSI 
\cite{Sun+09}. New $\beta$-decay halflives that impact weak and main $r$-process simulations have been measured at RIKEN 
\cite{Nishimura+11,Nishimura+12,Lorusso+15}, the NSCL \cite{Pereira+09,Hosmer+10,Quinn+12}, HRIBF 
\cite{Madurga+12,Mazzocchi+13,Miernik+13}, GSI \cite{Benzoni+12,Morales+14a,Morales+14c,Kurtukian-Nieto+14} and CERN/ISOLDE 
\cite{Arndt+11}. Neutron capture rates are inaccessible to direct measurements, however innovative indirect techniques are under 
development and show considerable promise \cite{Jones+11,Kozub+12,Spyrou+14}. Next-generation facilities such as FRIB, ARIEL, RIBF, 
SPIRAL2, ISOLDE upgrade, RIBLL, and FAIR will extend experimental reach by hundreds of more exotic isotopes. A pressing question 
therefore emerges---which of the thousands of isotopes potentially involved in an $r$-process are most important to measure?

One approach to determine which nuclear properties have the greatest astrophysical impact is to directly examine their influence on $r$-process 
abundance predictions via sensitivity studies. In an $r$-process sensitivity study, a baseline astrophysical trajectory is chosen and then run 
thousands of times, each with one piece of nuclear data systematically varied.  Results of the simulations are then compared to the baseline 
simulation with no data changes, to highlight the nuclei whose properties have the greatest leverage on the final abundance pattern in that 
particular environment. Sensitivity studies for a variety of main $r$-process conditions have been performed for nuclear masses 
\cite{Brett+12,Aprahamian+14,Surman+14b,Surman+14c,Mumpower+15a}, neutron capture rates \cite{Beun+09,Surman+09,Mumpower+12c,Surman+14c}, 
$\beta$-decay rates \cite{Surman+14c,Mumpower+14}, and are in progress for $\beta$-delayed neutron emission probabilities \cite{Surman+15}. 
Sensitivity studies of the weak $r$ process \cite{Surman+14a} and the related $i$ process \cite{Bertolli+13} also appear in the literature. The 
individual fission properties that are crucial for a fission recycling $r$ process have not yet been subject to the analogous sensitivity 
analysis; see \cite{Goriely15} for a recent review of fission models and their impact on a potential neutron star merger $r$-process abundance 
distribution.

Here we review the $r$-process sensitivity studies performed to date for a main $r$ process. We begin with a brief review of astrophysical 
environments and $r$-process dynamics in Section \ref{sec:astro} and a discussion of nuclear data inputs for $r$-process nuclear network codes in 
Section \ref{sec:nuc}. Global Monte Carlo estimates of the current predictability of $r$-process simulations are discussed in Section 
\ref{sec:monte}. We then present compilations of main $r$-process sensitivity study results for masses, neutron capture rates, $\beta$-decay 
rates, and $\beta$-delayed neutron emission probabilities in Section \ref{sec:sens}. We conclude with the experimental outlook, Section \ref{sec:exp}.

\section{Astrophysical environments}\label{sec:astro}

Heavy element abundance patterns roughly matching the solar $r$-process residuals have now been found in many metal-poor halo stars \cite{Sneden+08,Roederer+14}. 
The potential astrophysical site that perhaps most comfortably fits galactic chemical evolution models based on these observations \cite{Argast+04,Komiya+14,Matteucci+15} is within a core-collapse supernova. 
The supernova site that has received the most attention is the neutrino-driven wind off of the newly-formed protoneutron star (PNS) \cite{Meyer+92,Woosley+94,Qian:1996xt, Sumiyoshi:1999rh, Thompson:2001ys, Arcones:2010nj,Brett+12}. 
Here, the copious neutrino emission heats the material above the PNS and sets its composition. 
A combination of moderate neutron richness, high entropy, and fast outflow are required for the nucleosynthesis to reach uranium and thorium---conditions that are not achieved in modern simulations \cite{Fischer+10,Hudepohl+10,Roberts+12}.  
However, they are not terribly far off, sometimes reaching the second peak of the $r$ process \cite{Roberts:2011yw,MartinezPinedo:2012rb}. 
They will become more neutron rich with the inclusion of new physics, such as sterile neutrino \cite{ McLaughlin:1999pd,Beun:2006ka} of the type that is suggested by LSND \cite{Athanassopoulos:1996jb}, MiniBooNE \cite{AguilarArevalo:2010wv} and reactor neutron anomaly \cite{Hayes:2013wra}. 
Independently of new physics, the neutrino driven wind site deserves more study, as the theoretical error bar on the electron fraction has not been studied.

\begin{figure}
 \begin{center}
  \includegraphics[width=\textwidth]{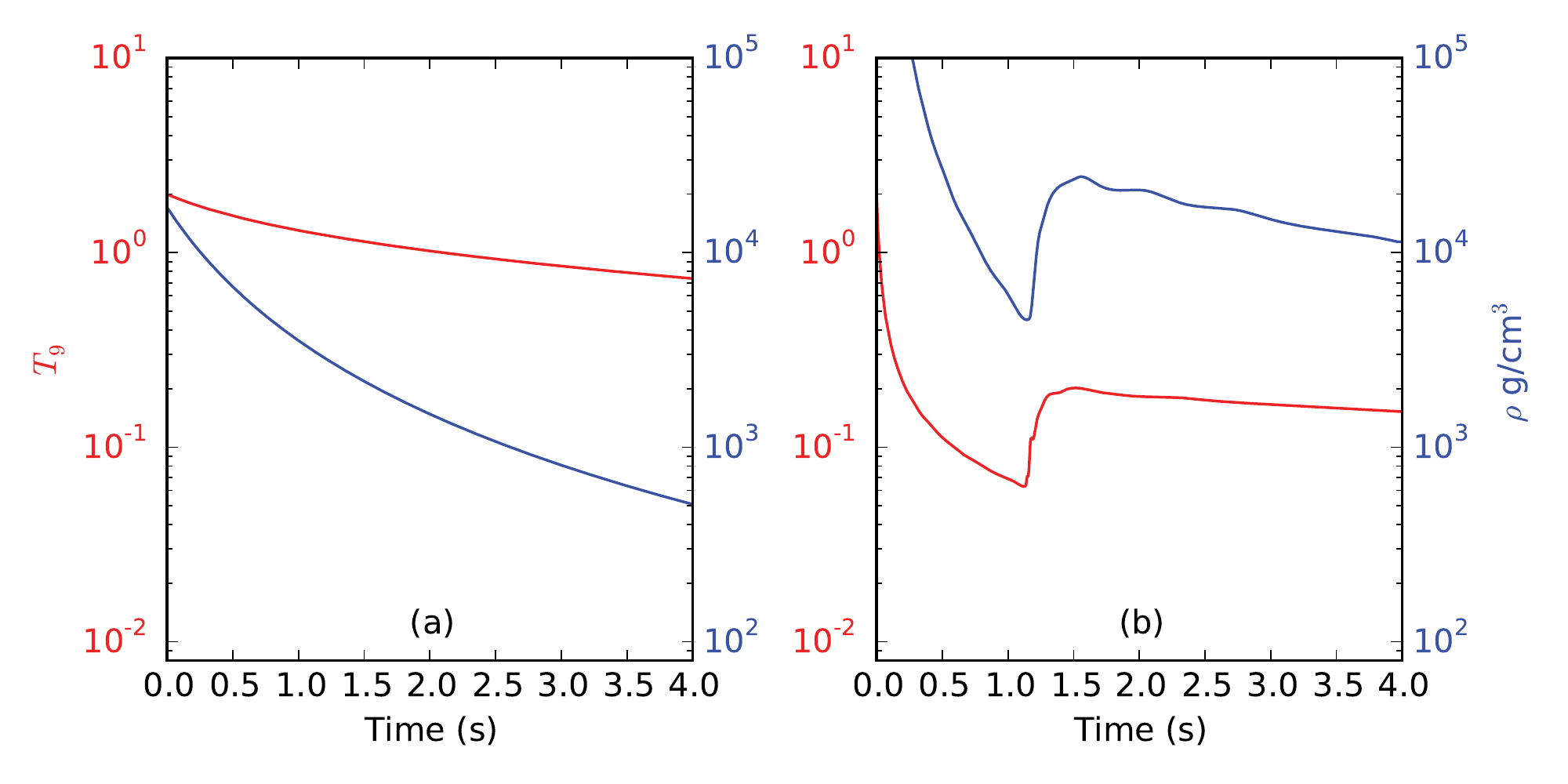}
  \caption{\label{fig:t9rho} Evolution of temperature and density as a function of time for a hot (a) and cold (b) $r$-process simulation. }
 \end{center}
\end{figure}

\begin{figure}
 \begin{center}
  \includegraphics[width=\textwidth]{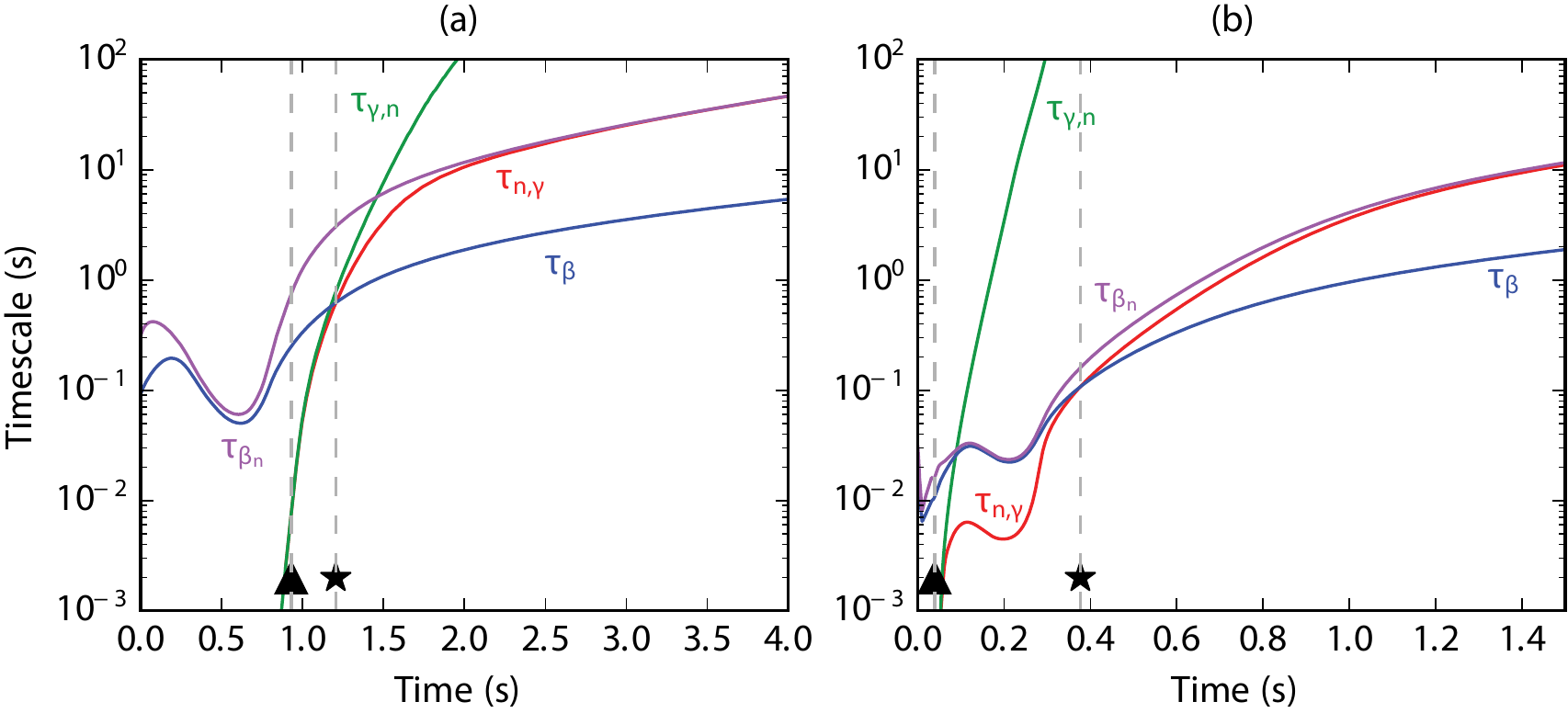}
  \caption{\label{fig:timescales} Abundance weighted timescales of important channels during the freeze-out phase for a hot (a) and cold (b) $r$ process. Two important points in the calculations are shown: break from \nggn \ equilibrium (triangle), and the point which $\beta$-decays take over neutron captures (star).}
 \end{center}
\end{figure}

\begin{figure}
 \begin{center}
  \includegraphics[width=\textwidth]{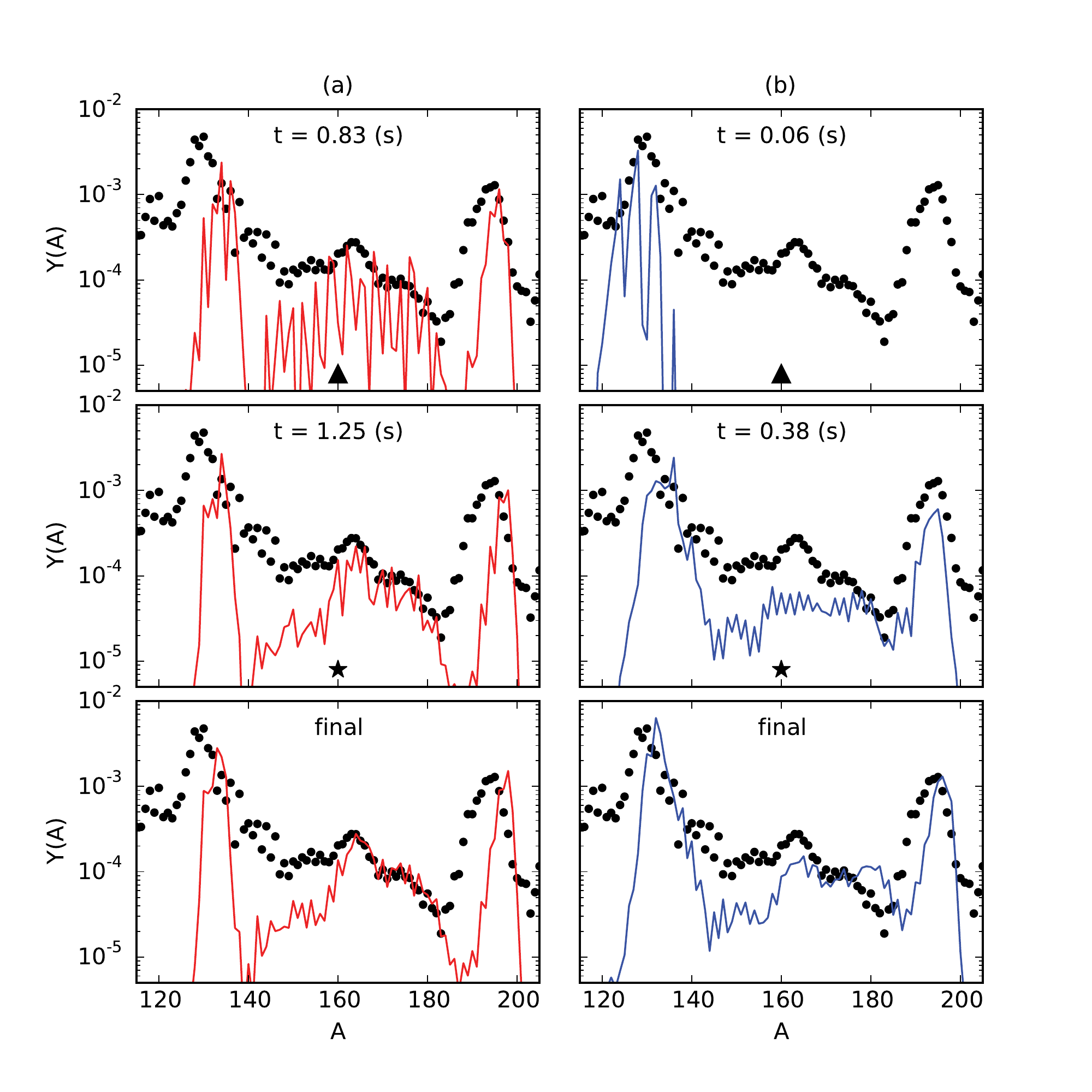}
  \caption{\label{fig:ab-evo} Evolution of isotopic abundances in hot (a) and cold (b) $r$-process simulations. Snapshots of abundances are 
shown at the two critical phase of Fig.~\ref{fig:timescales} along with the final pattern. }
 \end{center}
\end{figure}

The neutrino driven wind is an example of a warm (or `hot') environment. 
This type of $r$ process usually begins in an \nggn \ equilibrium, and stays in this equilibrium until the available neutrons are exhausted or almost exhausted. 
The path of the $r$ process depends primarily on the neutron separation energies of the various isotopes of each element and any type of structure that can be seen in the relative separation energies can leave its imprint on the final abundance pattern produced in these scenarios. 
However, a neutrino driven wind, if it accelerates quickly enough, can also be partially `cold', meaning that the system falls out of \nggn \ equilibrium before the neutrons are used up, and so, at least in the last moments, photodissociation plays little role in the nuclear flow that occurs during $r$-process nucleosynthesis. 
Instead it is a combination of $\beta$ decay and neutron capture that determine the path of the $r$ process. 
Thus, even within the neutrino driven wind environment, there exist different paths and different types of conditions that emphasize various aspects of the nuclei along the path. 
These different types of environments are illustrated in Figs.~\ref{fig:t9rho}, \ref{fig:timescales}, and \ref{fig:ab-evo}. 

Fig.~\ref{fig:t9rho} shows a sample evolution of temperature and density for a hot (left panel) and cold (right panel) mass trajectory. The 
hot $r$-process example is a wind parameterized as in \cite{Meyer+02} with entropy $30$ $k_B$, an initial electron fraction of $Y_e=0.20$, and 
a timescale of $70$ ms. The cold $r$ process is from the neutrino-driven wind simulations of Ref.~\cite{Arcones+07}, with the electron 
fraction reduced to $Y_e=0.31$ to produce a main $r$ process. Fig.~\ref{fig:timescales} shows the abundance weighted timescales for the 
relevant reactions during the nucleosynthesis that occurs using these trajectories. The abundance weighted timescales for neutron capture, 
photodissociation, total $\beta$ decay (with and without neutron emission), and $\beta$-delayed neutron emission shown in this figure are defined as 
follows:
\begin{subequations}\label{eqn:taus}
 \begin{equation}
 \label{eqn:tau-ng} \tau_{n,\gamma} = \frac{\sum_{Z\geqslant8,A}Y(Z,A)}{\sum_{Z\geqslant8,A}N_{n}\langle \sigma v\rangle_{Z,A} Y(Z,A)} 
 \end{equation}
 \begin{equation}
 \label{eqn:tau-gn} \tau_{\gamma, n} = \frac{\sum_{Z\geqslant8,A}Y(Z,A)}{\sum_{Z\geqslant8,A}\lambda_{\gamma n}(Z,A)Y(Z,A)} \\
 \end{equation}
 \begin{equation}
 \label{eqn:tau-beta} \tau_{\beta} = \frac{\sum_{Z\geqslant8,A}Y(Z,A)}{\sum_{Z\geqslant8,A}\lambda_{\beta}(Z,A)Y(Z,A)}
 \end{equation}
 \begin{equation}
 \label{eqn:tau-betajn} \tau_{\beta_{jn}} = \frac{\sum_{Z\geqslant8,A}Y(Z,A)}{\sum_{Z\geqslant8,A}\lambda_{\beta_{jn}}(Z,A)Y(Z,A)}
 \end{equation}
 \end{subequations}
In these equations the neutron number density is $N_{n}$, the neutron capture rate is $\lambda_{n \gamma} = N_{n}\langle \sigma 
v\rangle_{Z,A}$, the photodissociation rate is $\lambda_{\gamma n}$, the total $\beta$-decay rate is $\lambda_{\beta}$, and the sum of rates 
for $\beta$ decay followed by the emission of $j\geq1$ neutrons is $\lambda_{\beta_{jn}}$, for a nucleus with $Z$ protons and $A$ nucleons.

The left panel of Fig.~\ref{fig:timescales} shows these timescales for the hot $r$-process example whereas the right panel shows the 
timescales for the cold $r$-process example. In both figures the $(n,\gamma)$ and $(\gamma,n)$ lines (green and red, respectively) start out 
the same, indicating \nggn \ equilibrium. When the lines are no longer identical, equilibrium is lost, and this point is indicated by a 
triangle in each figure. In the cold trajectory, the lines diverge almost immediately. All $r$ processes have a phase where $\beta$-decay 
eventually takes over as the most important timescale for nuclear flow and this is indicated by a star in the figures. In hot scenarios 
$\beta$ decay becomes dominant just after \nggn \ equilibrium fails, during a period where neutron capture, photodissociation, and $\beta$ 
decay all compete. In cold scenarios $\beta$ decay takes over well after \nggn \ equilibrium has broken down and photodissociation ceases to 
play a role, and the dynamics are driven largely by the competition between $\beta$ decay/$\beta$-delayed neutron emission and neutron 
capture. Finally, both trajectories show that there is a period where neutron capture and $\beta$-delayed neutron emission timescales are 
closely matched. In the hot trajectory this occurs only at late times where relatively few neutrons are being captured, while in the cold 
trajectory these timescales are the most important ones for the bulk of the $r$ process \cite{Wanajo:2007jj}.

Fig.~\ref{fig:ab-evo} shows snapshots of the abundance pattern at various points in the calculation of the nucleosynthesis from Figs. 
\ref{fig:t9rho} and \ref{fig:timescales}. Again the left panel shows the hot trajectory and the right panel shows the cold trajectory. In the 
panels labeled with triangles, the abundance patterns are shown for the points in the simulation where the system is falling out of \nggn \ 
equilibrium, $\tau_{\gamma,n}/\tau_{n,\gamma}\gtrsim1$. A common feature of cold scenarios that is clearly shown in this figure is that 
relatively little neutron capture has occurred by the time the system drops out of equilibrium. The hot scenarios, on the other hand, show a 
fully populated abundance pattern. It is characterized by large odd-even staggering, since all of the equilibrium $r$-process waiting points 
are even-$N$ nuclei. The panels labeled by stars show abundance patterns at the time when $\beta$ decay takes over as the fastest timescale, 
$\tau_{\beta}/\tau_{n,\gamma}\sim1$. In the cold case, enough neutron capture has taken place that the pattern has stretched out to 
high mass number. In the hot case, some of the even-odd staggering has been smoothed out and the rare earth peak has begun to form. In both 
scenarios, however, the abundances are not finalized at this point. The abundance patterns change greatly during the decay back towards 
stability, as indicated by the final abundances shown in the bottom panels: the main peaks shift and widen or narrow, the rare earth peak is 
finalized, and the odd-even staggering continues to smooth out. Thus during the final stages of the $r$ process, individual nuclear 
properties may play a substantial role in the determination of final abundances.

An alternate, more robustly neutron-rich potential $r$-process site is within an neutron star-neutron star or neutron star-black hole merger 
\cite{Lattimer+74}. Modern simulations of the cold or mildly-heated merger tidal tail ejecta show a vigorous $r$ process with fission 
recycling \cite{Goriely+11,Korobkin+12,Just+15}. Fission recycling can produce an $r$-process abundance pattern between the second and third 
$r$-process peaks that is relatively insensitive to variations in the initial conditions \cite{Beun:2007wf}. This appears consistent with 
observations of the $r$-process-enhanced halo stars for which we have relatively complete $56<Z<82$ patterns---most are strikingly similar 
and a good match to the solar $r$-process residuals within this element range \cite{Sneden+08,Roederer+14}. If robust production of 
radioactive $r$-process species accompanies a merger, this will lead to an observable electromagnetic transient 
\cite{Li:1998bw,Metzger:2010sy,Roberts:2011xz,Barnes:2013wka}; one such event has perhaps already been discovered 
\cite{Tanvir:2013pia,Berger:2013wna}. Finally, the idea that neutron star mergers could be the origin of all $r$-process elements seems to 
fit some new numerical galactic simulations \cite{Shen:2013wva,vandeVoort:2014gca}, though the delay time for mergers to begin contributing 
to galactic nucleosynthesis is still uncertain \cite{Wanderman+15}.

\begin{figure}
 \begin{center}
  \includegraphics[width=\textwidth]{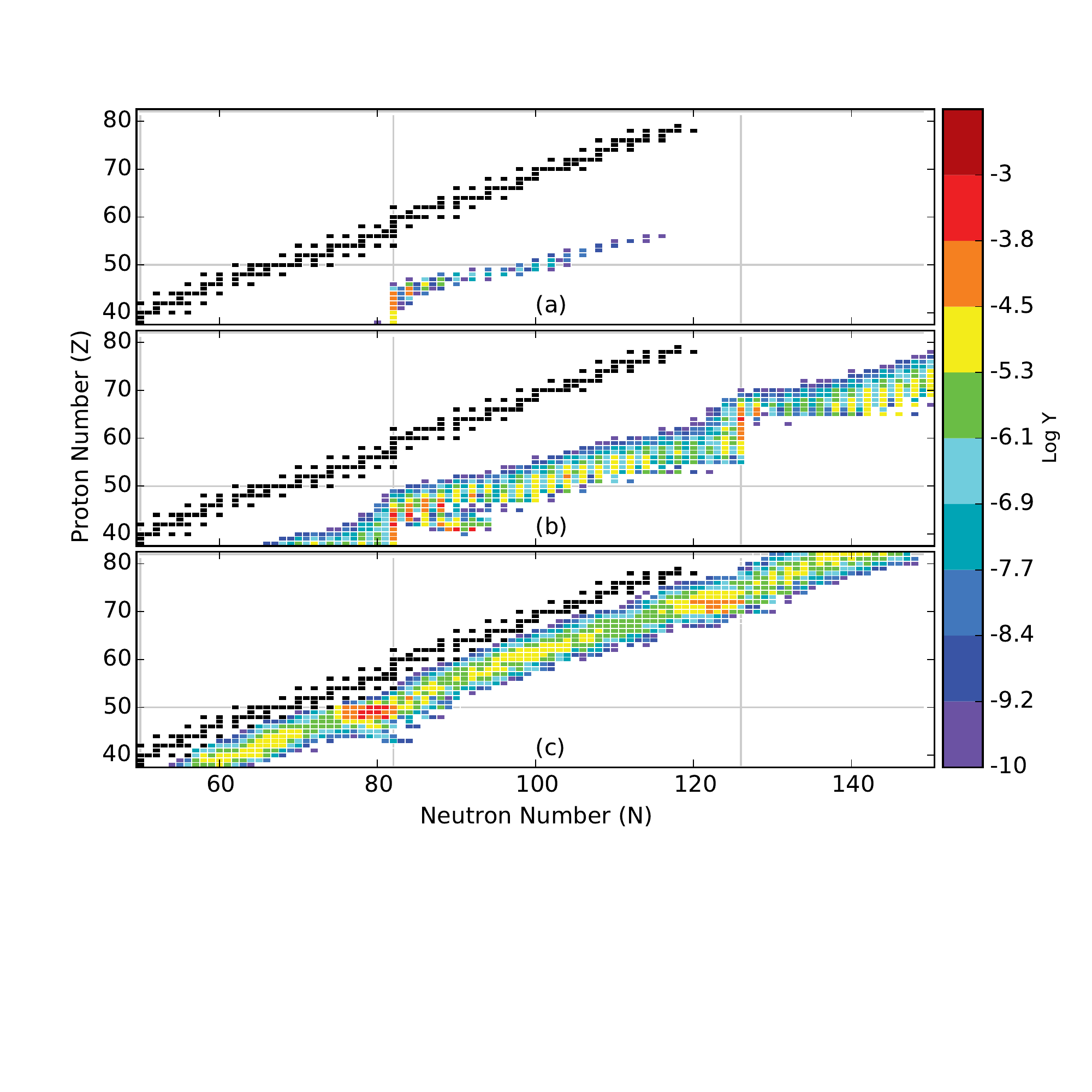}
  \caption{\label{fig:nsm-evo} The evolution of nuclear flow in a neutron star merger $r$-process. The simulation uses a neutron star merger 
trajectory from Bauswain and Janka as in Ref.~\cite{Goriely+11} and is calculated including fission recycling as described in 
Ref.~\cite{Surman+14c}. Snapshots taken at (a) the first hundred milliseconds (b) neutron exhaustion (neutron-to-seed ratio equal to unity) 
and (c) decay back to stability.}
 \end{center}
\end{figure}

The bulk of the $r$-process nucleosynthesis occurs from relatively cold material ejected in the tidal tails of the mergers. While the 
material can become heated sufficiently to have a brief \nggn \ equilibrium phase, for much of the process, the material is cold and it is 
the competition of neutron capture and $\beta$ decay that dominate. This environment differs from a cold neutrino driven wind trajectory in that 
it experiences a different density and temperature history and also in that it is much more neutron rich. Fig.~\ref{fig:nsm-evo} shows 
the path of the $r$ process at different times during a neutron star merger calculation. In fact, the material is so neutron rich that the 
path reaches the neutron drip line, as can be seen in panel (b) of Fig.~\ref{fig:nsm-evo}, and undergoes vigorous fission recycing.
 
There are two additional/alternative sources for $r$-process material in these objects. One arises from the disintegration of the disk at 
late times \cite{Metzger:2010sy} and the other comes from wind like outflows from the disk \cite{McLaughlin:2004be,Caballero:2011dw} or 
hypermassive neutron star \cite{Perego+14}. Particularly the latter environment would be more similar to that of a neutrino driven wind in 
the supernovae.
  
Neither the neutrino driven wind of supernovae nor neutron star mergers is a perfect fit to the available data. Thus, there is considerable 
work on alternative astrophysical sites for the $r$ process. For example, neutron rich jets from supernovae, e.g. \cite{Winteler+12}, are 
never heated to the higher entropies and $Y_e$s that typically obtain in the neutrino driven wind because the material accelerates away from 
the neutrinos too quickly. Thus the material retains more neutron-richness and can undergo an $r$ process.

Another example is neutrino-induced nucleosynthesis in the helium shell \cite{Banerjee+11} of core collapse supernovae. In this scenario, the 
neutrinos spall neutrons from pre-existing nuclei in the outer layers of the supernovae. The number of neutrons that are produced in this 
environment is strongly dependent on the energies of the incoming neutrinos since the cross section for spallation is highly energy 
dependent. This process produces neutrons at rate somewhat lower than typically considered in the $r$-process environments, although not as 
slowly as occurs in the $s$ process. Thus, much of the neutron capture occurs close to stability and the pattern it produces has some 
$s$-process, as well as $r$-process, characteristics.

A third example is accretion disk outflows from collapsars \cite{Surman+06}. This has many of the properties that one expects from the 
neutrino driven wind, except that the environment is expected to be considerably more neutron rich. There is a unique type of oscillation, 
the matter-neutrino resonance \cite{Malkus+12}, that can occur close to the neutrino decoupling surface that can cause a transition of 
electron type neutrinos into muon type neutrinos. The loss of these neutrinos allows the material to preserve some of its neutron richness 
from the disk, creating potentially favorable conditions for a rapid neutron capture process.

Five astrophysical trajectories that probe a range of $r$ process conditions are considered in this review. 
The first trajectory, trajectory (a), is a hot $r$ process with low entropy, $30$ $k_B$, an initial electron fraction of $Y_e=0.20$ and a timescale of $70$ ms. 
Trajectory (b) is also a hot $r$ process with high entropy $100$ $k_B$, an initial electron fraction of $Y_e=0.25$ and a timescale of $80$ ms. 
The hot $r$-process conditions are produced using a parameterized wind model from Ref.~\cite{Meyer+02} and are not neutron rich enough to lead to fission recycling. 
Trajectory (c) is a cold $r$ process from the neutrino-driven wind simulations of Ref.~\cite{Arcones+07}. 
This trajectory has an artificially reduced electron fraction of $Y_e=0.31$ which yields a main $r$ process, $A>120$. 
Trajectory (d) is a neutron star-neutron star merger from Bauswain and Janka as in Ref.~\cite{Goriely+11}. 
This trajectory is highly neutron rich and undergoes several fission cycles before the $r$ process ends. 
Trajectories (a) and (c) are used for the hot and cold $r$-process calculations of Figs.~\ref{fig:t9rho}, \ref{fig:timescales}, \& \ref{fig:ab-evo} and trajectory (d) is used in Fig.~\ref{fig:nsm-evo}. 
The sensitivity studies of Section \ref{sec:sens} use all four of these trajectories. 
A separate hot $r$-process trajectory with high entropy of $200$ $k_B$, an initial electron fraction of $Y_e=0.30$ and a timescale of $80$ ms is considered for the Monte Carlo studies of Section \ref{sec:monte}.

\section{Theoretical nuclear inputs}\label{sec:nuc}

The post-processing nucleosynthesis calculations described in the previous section require nuclear properties and reaction rates for thousands of nuclei from stability to the neutron drip line. 
Arguably the most important nuclear data sets for the $r$ process are nuclear masses, $\beta$-decay properties, and neutron capture rates. 
Theoretical models of these quantities are on relatively sure footing close to stability, where experimental information is available. 
Toward the drip line, different theoretical approaches produce markedly different (and often divergent) predictions. 
Here we compare sets of nuclear inputs commonly used in $r$-process simulations. 
Our aim is to motivate the choices of nuclear data variations in the sensitivity studies reviewed in Sec.~\ref{sec:sens}. 
A thorough review of the nuclear physics behind these models is found in \cite{Arnould+07} and references therein.

\subsection{Nuclear masses}

Nuclear masses are of fundamental importance for almost all areas of nuclear physics. 
Decades of development have gone into building global mass models with strong predictive power for a variety of nuclear physics applications including nuclear astrophysics. 
Modern mass formulae can be broadly grouped into three types: hybrid macroscopic-microscopic approaches that blend a liquid-drop model with microscopic corrections for shell effects and pairing, fully microscopic approaches based on solutions to the Schr{\"o}dinger equation for an effective potential, and empirical models that fit experimental masses with large numbers of parameters. 
The wide variety of approaches to nuclear mass models are reviewed in \cite{Lunney+03}; mass model developments since this review are compared in, e.g., \cite{Pearson+06,Hirsch+08,Hirsch+10,Barbero+12,Hua+12,Sobiczewski+14a,Sobiczewski+14b,Goriely+14}.

For the $r$ process, masses are used in the calculations of all nuclear quantities of interest. 
Generally it is mass differences that appear---neutron separation energies $S_n(Z,N)= M(Z,N-1)-M(Z,N)+M_n$ in calculations of neutron capture rates and photodissociation rates and $\beta$-decay $Q$-values $Q_{\beta}=M(Z,N)-M(Z+1,N-1)$, where $M(Z,N)$ is the atomic mass of the nuclide $(Z,N)$ and $M_n$ is the mass of the neutron. 
The largest dependence on masses becomes evident through the photodissociation rates, $\lambda_\gamma(Z,N)$, typically calculated from neutron capture rates and masses using detailed balance:
\begin{equation}\label{eqn:photo}
\lambda_\gamma(Z,N) \propto T^{3/2} \exp\left[-{\frac{S_n(Z,N)}{kT}}\right] \langle \sigma v \rangle_{(Z,N-1)}
\end{equation}
where $T$ is the temperature, $\langle \sigma v \rangle_{(Z,N-1)}$ is the neutron capture rate of the neighboring nucleus and $k$ is Boltzmann's constant. 
The neutron separation energy appears in the exponential, suggesting great precision is required to achieve reliable $r$-process predictions, e.g., Sec.~\ref{sec:sens}.

\begin{figure}
 \begin{center}
  \includegraphics[width=170mm]{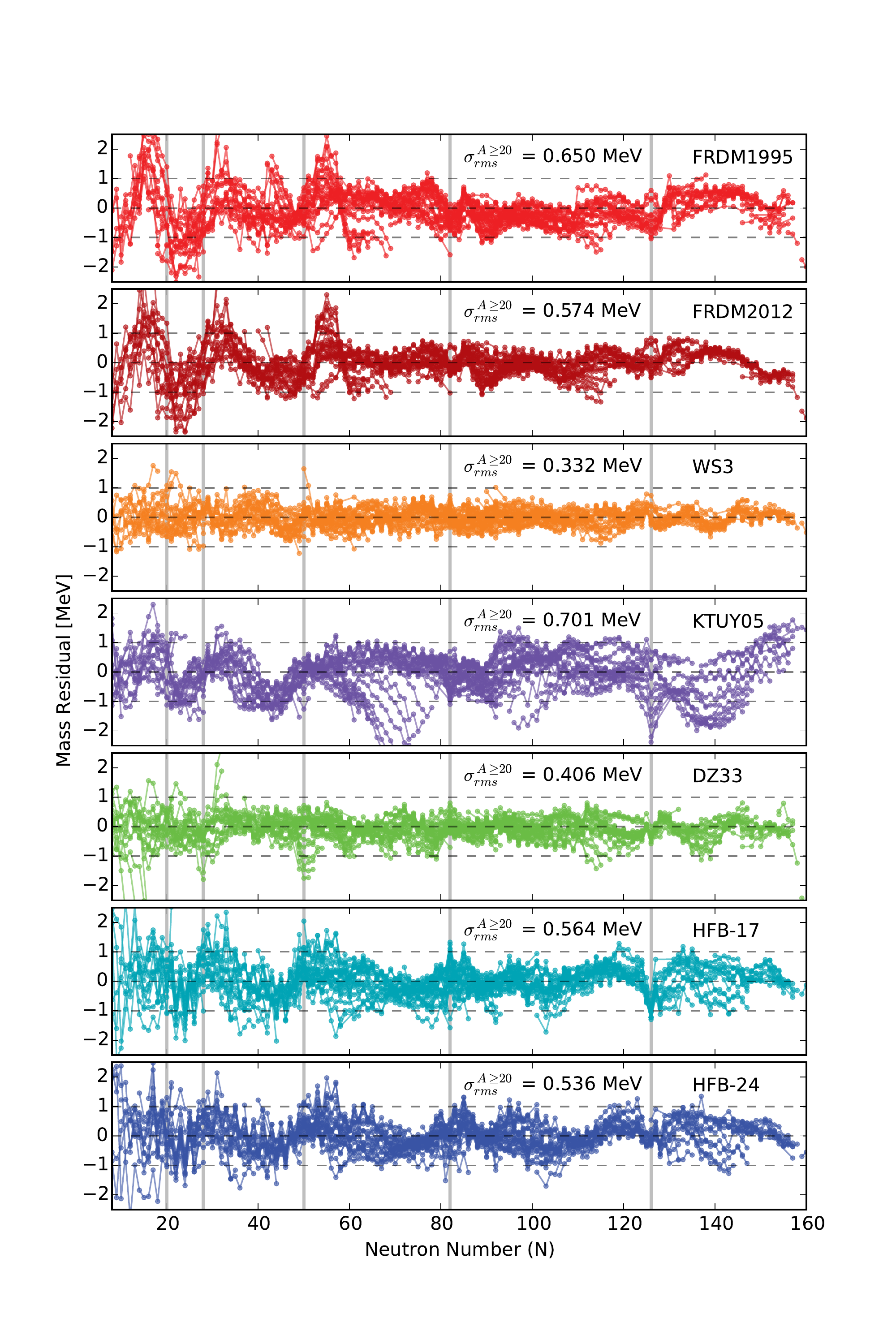}
  \caption{\label{fig:massexp} Predictions of seven commonly used mass models compared to the latest measured masses found in the 2012 Atomic Mass Evaluation (AME2012). 
  The rms errors shown here are quoted for nuclei with $A\geq20$.}
 \end{center}
\end{figure}

Figure \ref{fig:massexp} shows a comparison of the residuals resulting from taking the measured masses from the latest Atomic Mass Evaluation (AME2012) \cite{Audi+12} from the predicted masses of seven popular mass models. 
As much as $53\%$ new data exists between the latest evaluation (AME2012) and the previous evaluation (AME2003) \cite{Wapstra+03} most of which is from measurements on neutron-rich nuclei, thus providing more experimental input for $r$-process simulations. 
The Finite-Range Droplet Model (FRDM) \cite{Moller+95} is the \emph{de facto} standard for macroscopic-microscopic mass formulae. 
The macroscopic energy is calculated from a refined finite-range liquid-droplet model including a phenomenological exponential compressibility term, and the microscopic 
pieces include Strutinsky shell corrections calculated from a folded-Yukawa single-particle potential, pairing corrections evaluated using an effective-interaction pairing gap, and a Wigner term; a full description can be found in \cite{Moller+95,Lunney+03}. 
An updated version, FRDM2012, fit to AME2003 measured masses is newly available \cite{Kratz+14}; FRDM1995 and FRDM2012 are shown compared to AME2012 masses in the top two panels. 
FRDM2012 shows a significant decrease in rms value of nearly 100 keV relative to FRDM1995. 
The Weis{\"a}cker-Skyrme (WS3) model \cite{Liu+11,Zhang+14} is another macroscopic-microscopic approach, which aims for greater consistency of model parameters between the macroscopic and the microscopic parts. 
The latest version \cite{Zhang+14} is fit to AME2012 measured masses and is shown in the third panel. 
The Duflo-Zuker (DZ) \cite{Duflo+95,Kirson12} and KTUY05 \cite{Koura+05} are empirical formulae, directly fit to AME1995 \cite{Audi+95} and AME2003 measured masses, respectively, using slightly over 30 free parameters each. 
These models are shown compared to AME2012 in the fourth and fifth panels respectively. 
The leading example of a fully microscopic mass formula is based on the Hartree-Fock-Bogolyubov (HFB) approach, which has been used to produce a series of formulae, from HFB-1 \cite{Goriely+01} to HFB-24 \cite{Goriely+13b}, of increasing sophistication. 
The first version that achieved an rms deviation from experimental data below $0.6$ MeV is shown compared to AME2012 in the sixth panel, HFB-17 \cite{Goriely+09}. 
One of the latest version, HFB-24 \cite{Goriely+13b}, uses an extended Skyrme force with parameters (BSk24) fit to AME2012 measured masses and a delta-function pairing force derived from modern infinite nuclear matter calculations; its comparison with AME2012 is shown in the bottom panel. 
Despite their radically different approaches all of the models described above can reproduce measured masses within roughly $0.4$-$0.7$ MeV, as indicated in Fig.~\ref{fig:massexp}. 

\begin{figure}
 \begin{center}
  \includegraphics[width=\textwidth]{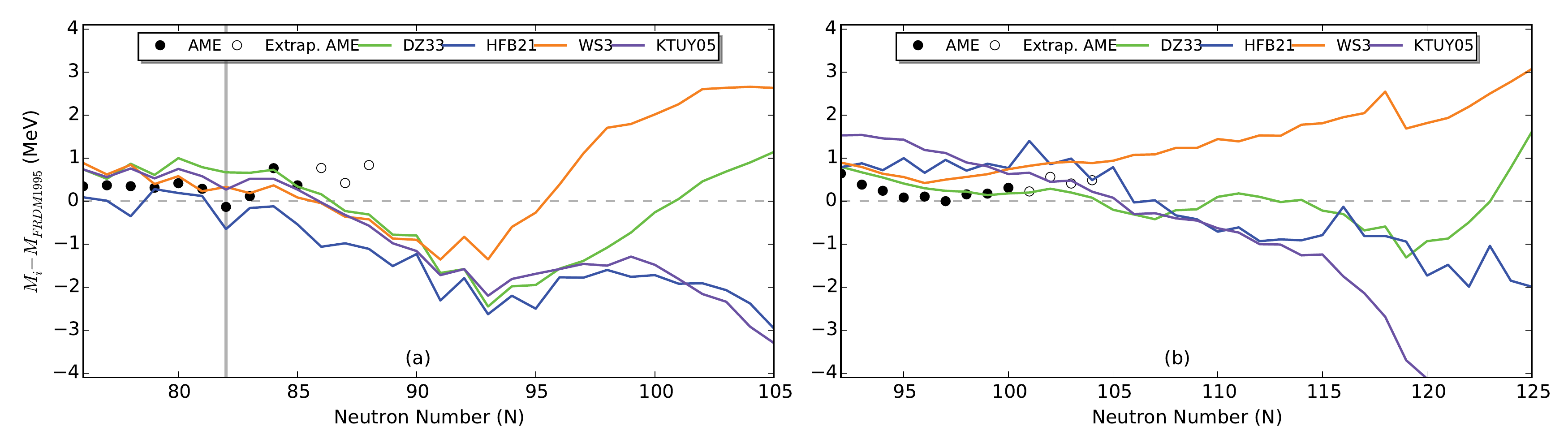}
  \caption{\label{fig:masstheo} Comparisons of measured measured masses to theoretical calculations for (a) Tin ($Z=50$, atomic symbol: Sn) and (b) Europium ($Z=63$, atomic symbol: Eu) isotopes. }
 \end{center}
\end{figure}

For the $r$ process, most of the relevant masses have not yet been measured. In fact, only a small handful of the measured masses in the 
AME2012 directly impact $r$-process calculations. Therefore, for the $r$ process the rms error is not nearly as important as the 
predictive power of the mass model. There are a couple of ways to estimate the uncertainty in the predictions of various mass models. One 
is to compare predictions to masses that have been measured after the mass model was published. Another, more suitable for situations 
where no measurement is available, is to compare the predictions of the mass models to each other. In Ref. \cite{Sobiczewski+14b} the 
predictions of various mass models that were fit using the measured masses tabulated in AME2003 were compared with the new measured 
masses reported in AME2012. It was found that the accuracy of the fit to the masses in AME2003 was not necessarily correlated with 
predictive power of the model for the new masses. For discussions regarding possible fundamental limitations on the theoretical 
description of nuclear masses see Refs.~\cite{Bohigas+02,Barea+05,Olofsson+06}.

The true theoretical uncertainty of a mass model when extrapolated far from stability is difficult to estimate. However, we can compare 
the predictions of various mass models. As an example, we plot in Fig.~\ref{fig:masstheo} experimental and theoretical masses for tin 
and europium isotopes as compared to FRDM1995. We see that farther from stability, the predictions become more and more different - up 
to several MeV. Discrepancies can exceed 10 MeV, particularly for the heaviest nuclei above the $N=126$ closed shell. Thus predictions 
for $S_n$ and $Q_\beta$ must become less and less reliable for the increasingly neutron-rich isotopes most important for the $r$ 
process. However, for the Monte Carlo studies of Sec.~\ref{sec:monte} and the sensitivity studies described in Sec.~\ref{sec:sens} which 
examine the impact of mass variations we will optimistically consider variations on order of the rms errors of the models, $\pm0.5$ MeV.

We conclude this subsection with a stray observation from the left panel of Fig.~\ref{fig:masstheo}. 
Note that for the tin isotopes with $86<N<94$, FRDM1995 predicts significantly less binding compared to all other models. 
This systematic trend is present for other isotopic chains in this transition region between the $N=82$ closed shell and deformed rare earth regions. 
The underestimate of stability in this region leads to a `hole' in the $r$-process abundance pattern around $A\sim140$ for simulations that use nuclear data based on FRDM1995 \cite{Mumpower+15a}. 
This feature is corrected in FRDM2012 \cite{Kratz+14}, as shown, for example, in Fig. 3 of Ref.~\cite{Mumpower+15b}. 

\subsection{$\beta$-decay properties}

As is the case with nuclear masses, a small but increasing number of the $\beta$-decay lifetimes required for $r$-process simulations have been determined experimentally; $\beta$-delayed neutron emission probabilities are known for only a handful of the relevant nuclei. 
Global theoretical models or extrapolations are required for the remainder. 
Building a realistic model for $\beta$ decay applicable over the entire neutron-rich side of the chart of the nuclides is a challenging proposition. 
Calculations of $\beta$-decay halflives demand ground state properties for the parent and daughter nuclei, as well as the full set of single-particle levels below threshold in the daughter. 
The shell model calculations that can provide this information for light nuclei are intractable for the medium-to-heavy $r$-process species. 
The alternate approaches currently in widespread use in $r$-process calculations include gross theory \cite{Takahashi+69,Koyama+70,Takahashi71} and a microscopic-macroscopic application of the Quasiparticle Random Phase Approximation (QRPA) \cite{Moller+03}.

The formula for $\beta$-decay half-lives can be written as \cite{Moller+03}
\begin{equation}
\frac{1}{T_{1/2}}=\sum_{0\leq E_{i}\leq Q_{\beta}} S_{\beta}(E_{i}) f(Z,Q_{\beta}-E_{i}),
\label{eq:beta}
\end{equation}
where $S_{\beta}(E_{i})$ is the $\beta$-strength function, $Q_{\beta}$ is the maximum $\beta$-decay energy, and 
$f(Z,Q_{\beta}-E_{i})$ is the Fermi function. The probabilities for $\beta$-delayed neutron emission then are given 
schematically by
\begin{equation}
P_{n}=\frac{\sum_{S_{n}\leq E_{i}\leq Q_{\beta}} S_{\beta}(E_{i}) f(Z,Q_{\beta}-E_{i})}{\sum_{0\leq E_{i}\leq Q_{\beta}} S_{\beta}(E_{i}) f(Z,Q_{\beta}-E_{i})}.
\label{eq:bdne}
\end{equation}
The $\beta$-strength function contains the nuclear matrix elements for the Gamow-Teller, Fermi, and first-forbidden $\beta$-decay operators (and in principle operators of higher orders, though these are less important and typically neglected). 
In gross theory, the discrete energy levels are smoothed out and approximated by statistical functions that are normalized to the sum rules. 
In \cite{Moller+03}, $S_{\beta}(E_{i})$ is calculated in the QRPA using a folded-Yukawa single-particle potential to which pairing and Gamow-Teller residual interactions are added. 
The pairing interaction and effective potential are the same as employed in the FRDM, which is used to estimate the $Q_{\beta}$ values. 
For both models, first-forbidden contributions are estimated in gross theory. 
A more detailed summary of these calculations and a comparison with more sophisticated (and purely microscopic) $\beta$-decay estimates performed for limited sets of nuclei can be found in \cite{Arnould+07}.

\begin{figure}
 \begin{center}
  \includegraphics[width=\textwidth]{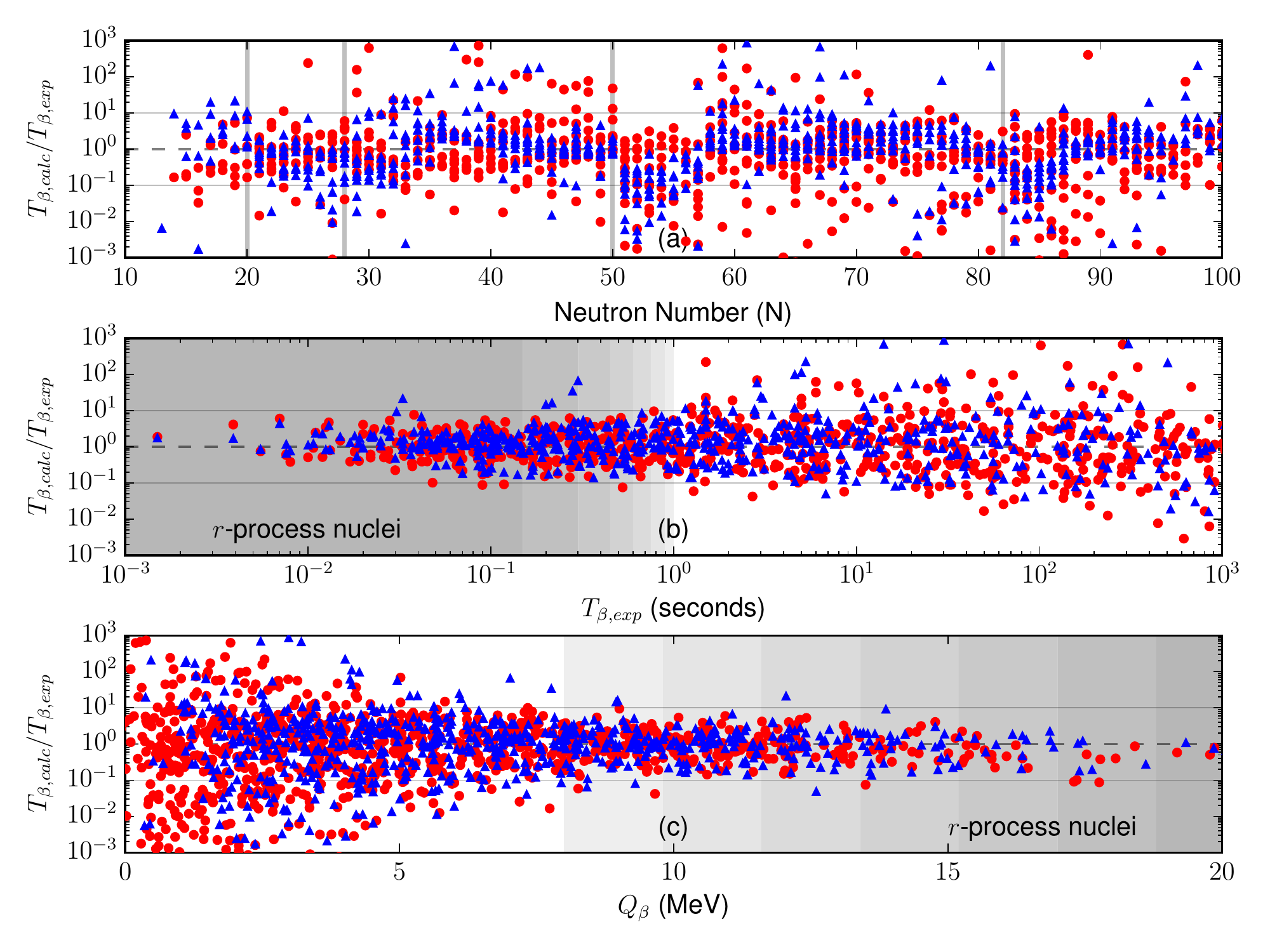}
  \caption{\label{fig:betaexp} Comparison of theoretical $\beta$-decay half-lives to measured values from the NNDC database \cite{NNDC} versus (a) neutron number, (b) measured half-life and (c) calculated $\beta$-decay Q-values. FRDM1995 + QRPA data points denoted by red circles and KTUY05 + gross theory data points denoted by blue triangles. }
 \end{center}
\end{figure}

Fig.~\ref{fig:betaexp} shows a comparison between experimental halflives as tabulated in the NNDC \cite{NNDC} $T_{\beta,exp}$ and theoretical values $T_{\beta,calc}$ from the global QRPA calculations of \cite{Moller+03} with FRDM1995 $Q_{\beta}$-values and the gross theory calculations of \cite{Nakata+97,Yoshida+00} with KTUY05 $Q_{\beta}$-values. 
The ratio $T_{\beta,calc}/T_{\beta,exp}$ is plotted three ways: versus neutron number in the top panel, versus $T_{\beta,exp}$ in the middle panel, and versus calculated $Q_{\beta}$-values in the bottom panel. 
The comparison in the top panel is at the first glance troubling, as the ratio $T_{\beta,calc}/T_{\beta,exp}$ spans six orders of magnitude. 
However, it is clear from the middle and bottom panels that the extreme discrepancies with theory are limited to the longest halflives and smallest $Q_{\beta}$-values. 
This is due to the phase space factor that appears in Eqn.~\ref{eq:beta}, which goes roughly as $(Q_{\beta}-E_{i})^{5}$. 
Decays with low $Q_{\beta}$s have a smaller `window' for energetically-allowed transitions, thus fewer transitions are available, and the difference $Q_{\beta}-E_{i}$ is much more sensitive to small variations in predicted transition energies. 
Most of the decays relevant for the $r$ process, on the other hand, have short halflives and large $Q_{\beta}$ values, and even the schematic gross theory can predict experimental 
values to within an order of magnitude for these nuclei. 
Thus, a factor of $10$ is chosen for the rate variations in the $\beta$-decay sensitivity studies described in Sec.~\ref{sec:sens}.

\subsection{Neutron capture rates}

While measurements are becoming available for an increasing number of nuclear masses and $\beta$-decay rates of interest for the $r$ process, direct measurements of neutron capture on unstable nuclei are not currently feasible (though see, e.g., \cite{Reifarth+14}). 
Presently all capture rates for the $r$ process are calculated via the Hauser-Feshbach (HF) statistical model \cite{Hauser+52}. 
In the HF model of $(n,\gamma)$ reactions, the captured neutron and target nucleus form a compound system that exists long enough to come into thermodynamic equilibrium, and then decays via $\gamma$ emission. 
The cross section for the reaction $I^{\mu} + n \rightarrow L + \gamma$, where the target nucleus $I$ is initially in state $\mu$, is given by
\begin{equation}
\sigma^{\mu}_{n,\gamma}(E)=\frac{\pi}{k^{2}\left(2J_{I}^{\mu}+1\right)\left(2J_{n}+1\right)}\sum_{J^{\pi}} (2J+1)\frac{T_{n}^{\mu}(J^{\pi})T_{\gamma}(J^{\pi})}{T_{tot}(J^{\pi})},
\label{eq:HF}
\end{equation}
where $k$ is the neutron wave number $k=\sqrt{2M_{In}E_{cm}}/\hbar$, with $M_{In}$ the reduced mass and $E_{cm}$ the center-of-mass energy; $J_{n}$ and $J_{I}^{\mu}$ are the spins of the neutron and target nucleus; $T_{tot}$ is the total transmission function for the decay of the compound nucleus; $T_{n}^{\mu}$ and $T_{\gamma}$ are the transmission functions for the formation and decay channels, respectively; and the sum is over all possible states $J^{\pi}$ in the compound nucleus. 
Ideally the transmission functions are calculated from experimental data; for $r$-process nuclei, however, little to no structure information is available. 
Models of nuclear level densities (for $T_{n}$) and $\gamma$-strength functions (for $T_{\gamma}$) are required in lieu of experimental data. 
Different choices for these quantities and the other inputs to HF codes---masses, deformations, particle optical potential models, treatment of a direct capture component, if any---lead to large variations in the capture rate predictions \cite{Arnould+07,Rauscher+12}.

\begin{figure}
 \begin{center}
  \includegraphics[width=\textwidth]{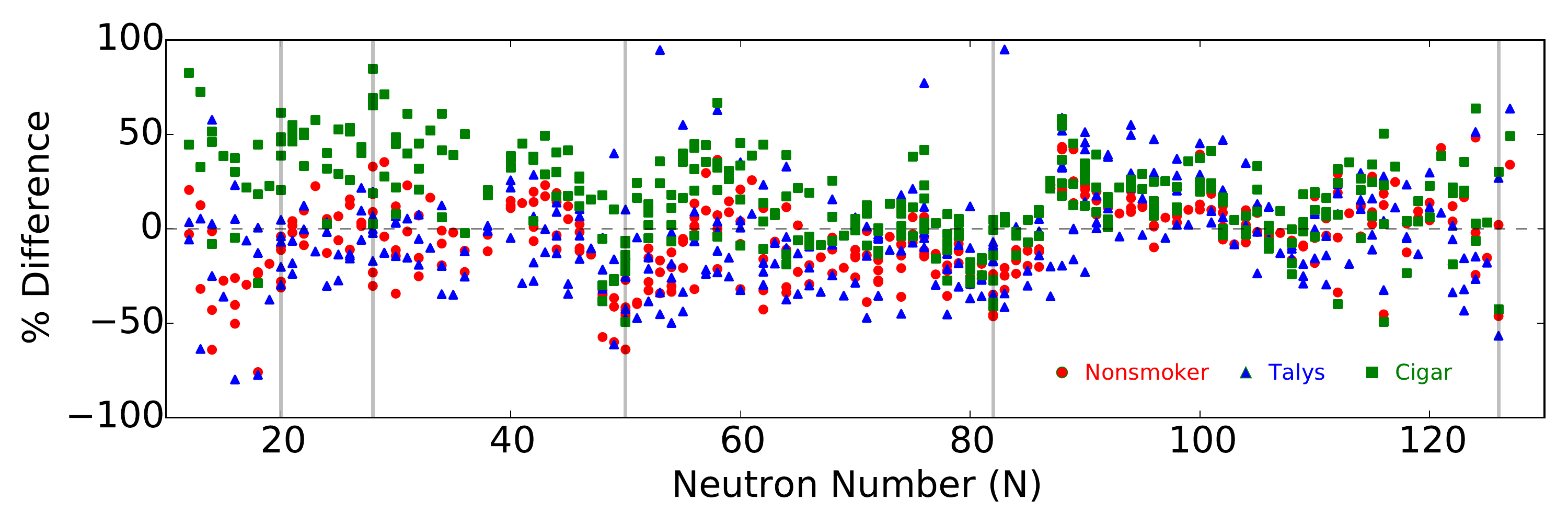}
  \caption{\label{fig:ncapexp} Comparison of Maxwellian averaged cross sections from various statistical model codes, Non-smoker \cite{NONSMOKER}, TALYS \cite{TALYS} and CIGAR \cite{Beard+14}, at $T_9=1.0$ to the KADoNiS database \cite{KADONIS}.}
 \end{center}
\end{figure}

Fig.~\ref{fig:ncapexp} compares the predictions of three HF calculations of Maxwellian-averaged neutron capture rate cross sections with the experimental values compiled in the KADoNiS database \cite{KADONIS}. 
The three model calculations include the widely-used NONSMOKER rates \cite{NONSMOKER} and rates calculated using the publicly-available TALYS code with microscopic input parameters consistent with HFB. 
The third set is from a newly-updated version of NONSMOKER, CIGAR \cite{Beard+14}. 
The KADoNiS database contains only nuclei on or very close to stability, so the transmission functions in Eqn.~\ref{eq:HF} can be estimated at least in part from experimentally-known levels. 
Even here, the predictions of different codes can vary by factors of two. 
A detailed comparison of this data is found in \cite{Beard+14}.

\begin{figure}
 \begin{center}
  \includegraphics[width=\textwidth]{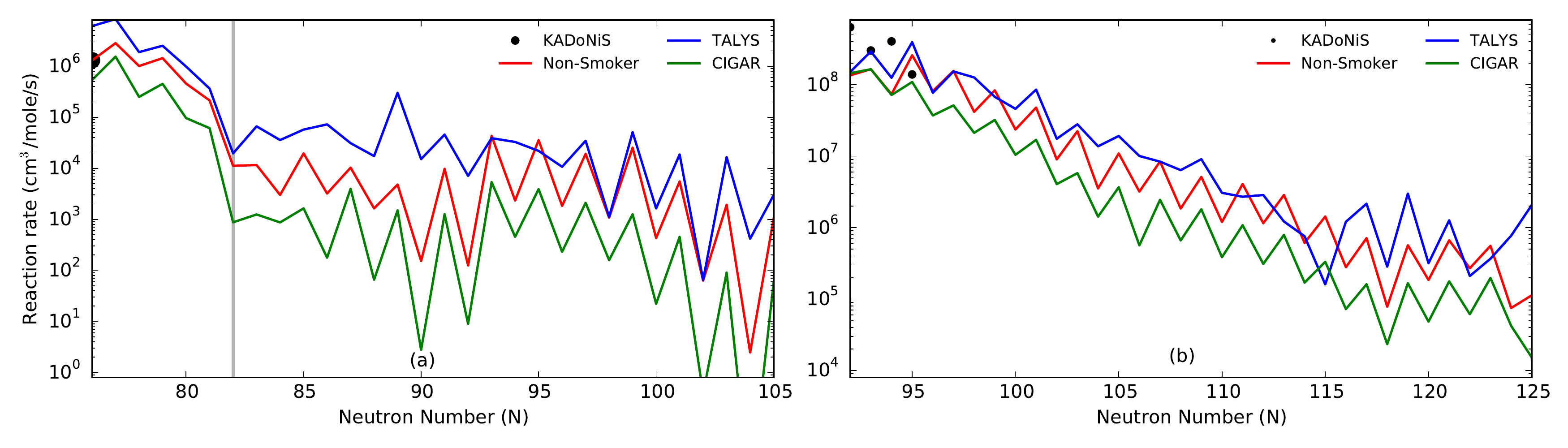}
  \caption{\label{fig:ncaptheo} Comparisons of KADoNiS reaction rate data at $T_9=1.0$ to theoretical calculations for (a) Tin ($Z=50$, atomic symbol: Sn) and (b) Europium ($Z=63$, atomic symbol: Eu) isotopes. }
 \end{center}
\end{figure}

Away from stability, where the rates are no longer constrained by experiment, the model variations are much wider. 
This is illustrated in Fig.~\ref{fig:ncaptheo}, which shows rate comparisons for the tin and europium isotopic chains from NONSMOKER and calculated with TALYS and CIGAR as described above. It is clear from this figure that there are no neutron capture rates in the KADoNiS database relevant to the $r$ process, and that the different model predictions can disagree by over three orders of magnitude. 
The neutron capture rate sensitivity studies described in Sec.~\ref{sec:sens} use an optimistic factor of $100$ for the rate variations.

\section{Monte Carlo variations of nuclear properties}\label{sec:monte}

Section~\ref{sec:nuc} presented rough estimates of the uncertainties in the theoretical nuclear masses, $\beta$-decay halflives, and 
neutron capture rates important for $r$-process simulations. We wish to understand how the uncertainties in these quantities 
propagates to uncertainties in the overall abundance pattern. To answer this in a quantitative manner, we examine Monte Carlo 
variations of nuclear properties as described in \cite{Mumpower2014_CGS15,Mumpower2014F}.

In the Monte Carlo approach, individual nuclear properties are varied throughout the nuclear chart using a probability distribution based on 
estimates of their theoretical uncertainties. For each set of varied nuclear inputs, an $r$-process simulation is repeated and a final 
abundance pattern generated. The ensemble of abundance patterns produced in this way is then analyzed statistically.

The choice of nuclear property to probe in a Monte Carlo study strongly impacts the choice of probability distribution. 
For nuclear masses, a normal distribution, 
\begin{equation}\label{eqn:mass_dist}
p(x) = \frac{1}{\sigma\sqrt{2\pi}} \exp\left[-\frac{(x-\mu)^2}{2\sigma^2}\right]
\end{equation}
can be used for each nucleus $(Z,A)$ in the simulation where the distribution is centered on the theoretical mass prediction, $\mu = M(Z,A)$, with $\sigma$ taken to be the rms value of the mass model and $x$ is the real-valued support of $p$.
The change to a given nuclear mass for the $i$-th Monte Carlo step is then
\begin{equation}\label{eqn:mass_delta}
M_i(Z,A) = M(Z,A) + \Delta_i(Z,A)
\end{equation}
where $\Delta_i(Z,A)$ is the random variable generated by sampling the normal distribution. 
At each Monte Carlo step a new mass table is generated from which new $Q$-values and other properties that depend on nuclear masses may be computed. 
Separate studies of reaction rates and half-lives can also be probed using this approach as well. 
However, in this case instead of generating a random \textit{additive} factor such as $\Delta$ for the nuclear masses, $p(x)$ is used to generate a \textit{multiplicative} factor to alter a given rate. 
The multiplicative factors are taken from a log-normal distribution,
\begin{equation}\label{eqn:rate_dist}
p(x) = \frac{1}{x\sqrt{2\pi}\sigma} \exp\left[-\frac{(\mu-ln(x))^2}{2\sigma^2}\right]
\end{equation}
where $\mu$ is the mean, and $\sigma$ is the standard deviation of the underlying normal distribution and $x$ is the support of $p$ for reals greater than 0. 

Global Monte Carlo studies of nuclear properties follow a fairly general implementation: 
\begin{enumerate} 
\item Fix choice of astrophysical conditions and nuclear model. 
\item Sample the probability distributions for all nuclei given the select nuclear property to study. 
\item Run an $r$-process simulation which encodes the impact of the changed nuclear physics. 
\item Repeat steps 2 and 3 until statistically significant iterations have been achieved. 
\item Calculate the variance in abundance patterns to estimate the error bars on final abundances. 
\end{enumerate}

\subsection{Uncertain nuclear masses}

\begin{figure}
 \begin{center}
  \includegraphics[width=\textwidth]{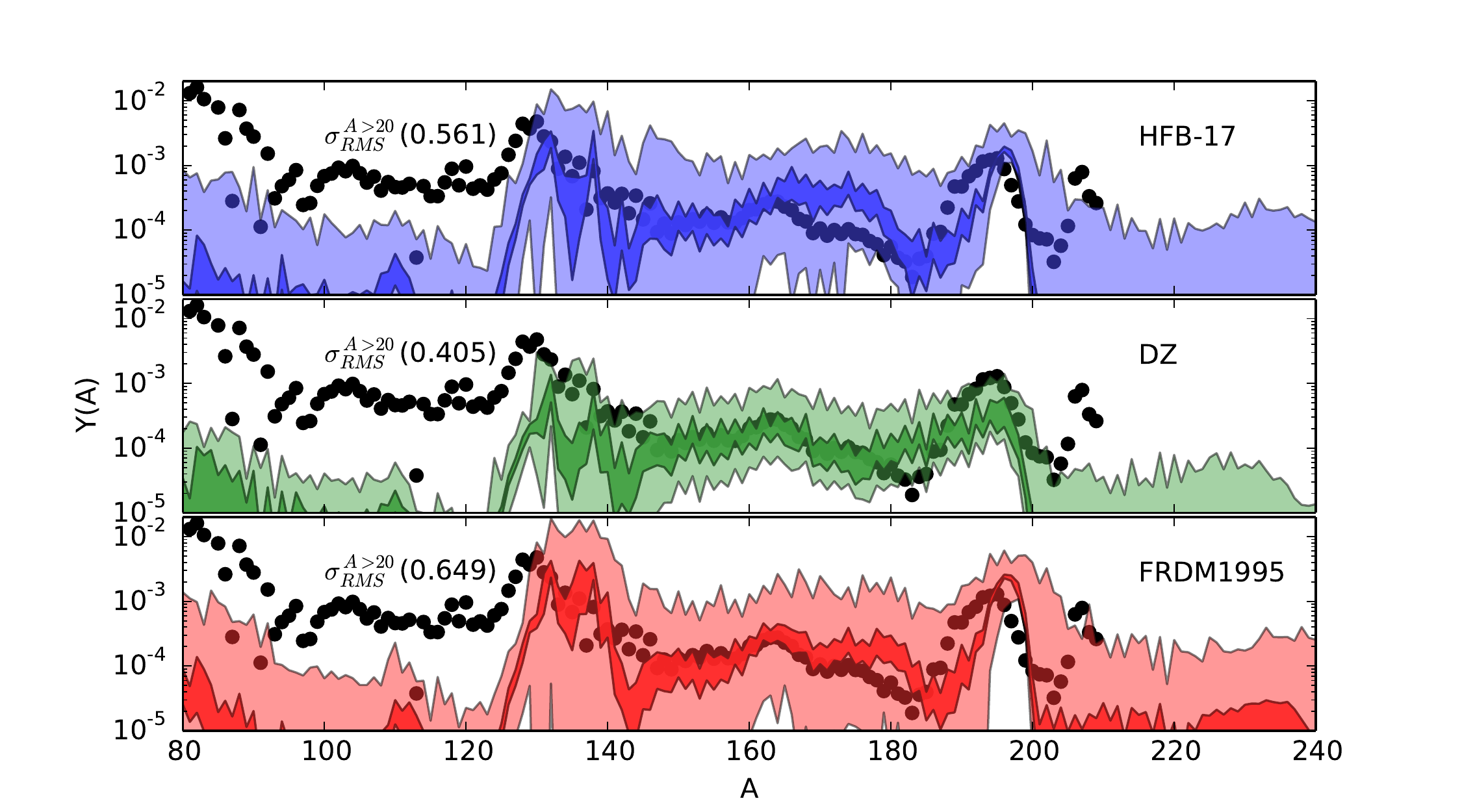}
  \caption{\label{fig:ab-err_mass} Variance in isotopic abundance patterns from three nuclear mass model predictions (HFB-17, DZ33, and FRDM1995) compared to the solar data (dots) for a hot $r$-process trajectory that produces primarily main ($A>120$) $r$-process nuclei. Darker shaded band represents Monte Carlo simulation with mass model rms error hypothetically reduced to $100$ keV. Simulation data from \cite{Mumpower2014_CGS15}. }
 \end{center}
\end{figure}

The error bars from such a study of uncertain nuclear masses in the context of a classical hot $r$-process is presented in Fig.~\ref{fig:ab-err_mass}. 
In this study from Ref. \cite{Mumpower2014_CGS15}, mass variations above and below theoretical values are sampled with equal weight using Eqn. \ref{eqn:mass_dist} with $\mu=0$. 
The size of the variation corresponds to the quoted rms error of the mass model when compared to measured nuclei, $\sigma\sim500$ keV. 
Thus this is an optimistic scenario. 

Under this assumption, the lighter shaded band represents the current predictive power of three leading mass models in Fig.~\ref{fig:ab-err_mass}. 
It is evident from this band that current mass model rms uncertainties have a large influence on the prediction of final abundances. 
Nearly all of the solar isotopic residual data for a main $r$ process lie within the abundance predictions of these models. 
A hypothetical reduction of each mass model rms error to $100$ keV is shown by the darker bands in Fig.~\ref{fig:ab-err_mass}. 
Abundance predictions between different mass models at this reduced level of uncertainty become distinct allowing one to clearly distinguish between the predictions of different nuclear models. 

In this study, the variation in nuclear masses is propagated to separation energies which go into the calculation of photodissociation rates. 
This has been shown to be a good approximation under hot astrophysical conditions \cite{Surman+09,Aprahamian+14}. 
Propagating changes in nuclear masses consistently to all nuclear properties that depend on masses as in \cite{Mumpower+15a,Mumpower+15b} will allow this method to be applied to other $r$-process conditions where photodissociation does not play such a prominent role. 
However, this is at the cost of greatly increasing the computational power needed for these studies as all properties of the given nuclear model must be recomputed at each Monte Carlo step. 

\subsection{Uncertain $\beta$-decay and neutron capture rates}

\begin{figure}
 \begin{center}
  \includegraphics[width=\textwidth]{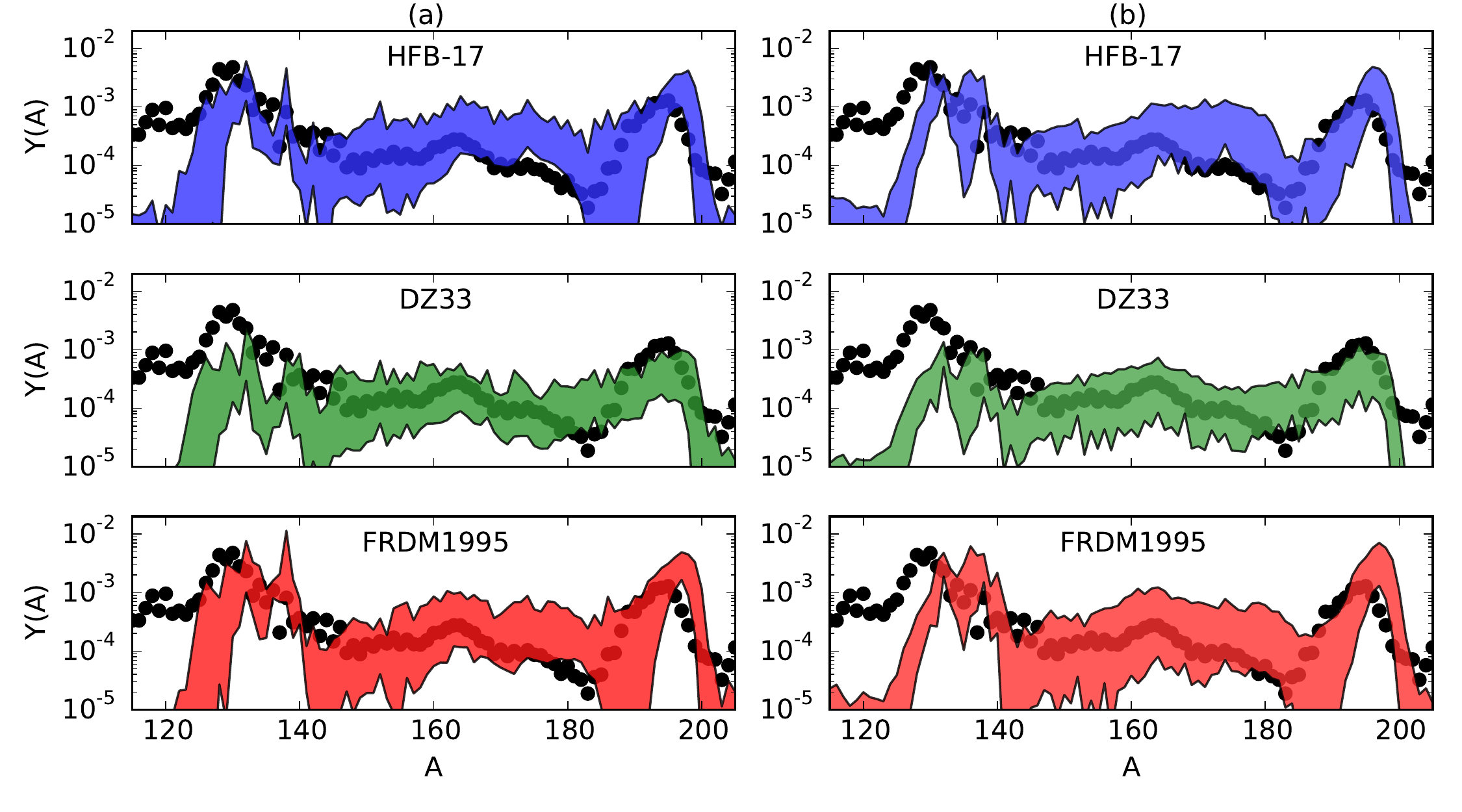}
  \caption{\label{fig:ab-err_rates} Variance in isotopic abundance patterns from uncertain $\beta$-decay half-lives, panel (a) and uncertain neutron capture rates, panel (b). The same three nuclear mass model predictions (HFB-17, DZ33, and FRDM1995) and the same main ($A>120$) $r$-process conditions are used as in Fig.~\ref{fig:ab-err_mass}. Simulation data from \cite{Mumpower2014F}. }
 \end{center}
\end{figure}

The variance bands in final abundances from uncertain rates is shown in Fig.~\ref{fig:ab-err_rates}. 
In these calculations, the same Monte Carlo approach has been applied separately to $\beta$-decay rates, panel (a), and neutron capture rates, panel (b) \cite{Mumpower2014F}. 
To approximate current theoretical uncertainties in $\beta$-decay and neutron capture rates, multiplicative factors are generated from log-normal distributions with underlying normal distribution values: $\mu=0$ and $\sigma=\text{ln}(2)$ and $\mu=0$ and $\sigma=\text{ln}(10)$ respectively. 
This yields multiplicative factors that range from $10^{-1}$ to $10$ for $\beta$ decays and $10^{-3}$ to $10^{3}$ for neutron capture rates, in agreement with the range of theoretical calculations of these quantities as can be seen in Figs.~\ref{fig:betaexp} and \ref{fig:ncaptheo}. 

Changes to $\beta$-decay rates can influence the pattern during the entire $r$-process evolution; the large variances shown in Fig.~\ref{fig:ab-err_rates} (a) are thus not too surprising. 
Neutron capture rates, however, influence the pattern only after the long duration \nggn \ equilibrium has ended in this environment, as illustrated in Fig.~\ref{fig:timescales}. 
Figure \ref{fig:ab-err_rates} (b) clearly shows that neutron captures play a critical role in the formation of the final abundances during the last moments of the $r$ process. 

The uncertainty in both these types of rates produces roughly the same order of magnitude variance in $r$-process abundances as from the uncertainty in nuclear masses. 
Taken together, the results of Figs. \ref{fig:ab-err_mass} and \ref{fig:ab-err_rates} imply that current error bars are too large to distinguish between the predictions of nuclear models. 
In order to improve the predictability of $r$-process simulations, advances in the description of neutron-rich nuclei must be achieved. 

\section{Sensitivity studies}\label{sec:sens}

As illustrated in the previous sections, present-day uncertainties in individual nuclear properties strongly limit the predictive power of 
$r$-process simulations. Improvements will require progress in modeling nuclei far from stability as well as new experiments that measure key 
quantities directly. Sensitivity studies are one way to identify these key quantities. In this section we focus on a particular type of 
sensitivity, which is the sensitivity of the properties of specific nuclei to the largest magnitude changes in the overall abundance pattern. 
In addition to this type of sensitivity, other, equally important sensitivities exist, which are the properties of specific nuclei, or groups 
of nuclei, to local structure in the abundance pattern, but these will be covered elsewhere.

In general, purpose of any type of sensitivity study is to gauge the astrophysical response of a change in nuclear physics input(s). The 
power of these studies is to point out the nuclear properties which play the most important role in shaping the final abundances observed in 
nature. Sensitivity studies thus play a key role in facilitating state-of-the-art measurements as they provide crucial astrophysical 
motivation to focus experimental campaigns on the most impactful nuclei.

A sensitivity study begins with a `baseline' simulation which defines the choice of astrophysical conditions and inputs from 
nuclear models. Subsequent simulations are then performed with this fixed input, but allowing a subset of the nuclear input 
data to vary. This enables the investigation of individual rates, masses, or branching ratios in the context of the given 
astrophysical conditions.

In a sensitivity study, each time nuclear data is changed, a corresponding final abundance pattern is produced. 
It is this one-to-one correspondence between varied nuclear data and abundance pattern that allows one to measure the astrophysical impact. 
Generally, one compares the baseline simulation to the simulation with varied nuclear data as with the impact parameter, $F$, defined as 
\begin{equation}\label{eqn:F}
F=100\sum_{A}|X(A)-X_{b}(A)|
\end{equation}
where $X_{b}(A)$ is the final isobaric mass fraction in the baseline simulation, $X(A)$ is the final isobaric mass fraction of the simulation with varied nuclear data input, and the summation runs over the entire baseline pattern. 
The isobaric mass fractions are defined by $X(A)\equiv\sum\limits_{Z+N=A}X(Z,N)$ where $X(Z,N)$ is the mass fraction of a species in the reaction network given $Z$ protons and $N$ neutrons. 
The total mass fraction always sums to unity and an isobaric mass fraction can be transformed into an abundance by the relation $X(A)=AY(A)$. 

It is important to keep in mind that the impact parameter, $F$, is a global sensitivity metric as it is defined as a difference in final abundances (or mass fractions) from two separate simulations. 
Thus, the metric is weighted in favor of nuclei with higher abundance. 
This can be seen by noting small differences between two large numbers will likely produce a larger $F$ than larger differences between two small numbers. 
Gauging the impact of local changes is best suited for metrics which reduce this preference, for example, by taking each component of the sum of Eqn. \ref{eqn:F} and dividing by the baseline mass fraction as in Ref. \cite{Mumpower+12c}. 

\subsection{Nuclear masses}

Nuclear masses are arguably one of the most important nuclear physics inputs that go into simulations of the $r$ process as they enter into the calculations of all other relevant nuclear physics quantities \cite{Burbidge+57,Cameron57}. 
This includes, for instance, the calculations of particle thresholds, reaction rates, half-lives and branching ratios \cite{Mumpower+15a}. 

Nuclear masses enter into the calculation of $\beta$-decay half-lives mainly via the phase space component which depends roughly on the fifth power of the $\beta$-decay energy to a given state for allowed decays. 
To a lesser extent nuclear masses impact the nuclear matrix elements of QRPA calculations \cite{Fang+13}. 
Nuclear masses enter into the calculation of neutron capture rates primarily via the $\gamma$-strength function, level density, and particle optical model. 
The choice of Hauser-Feshbach (HF) model inputs complicates the dependencies on nuclear masses and differences between masses. 
Masses can appear in several places, for instance, in the constant temperature term, shell correction term, the level density parameter or in the definition of the nuclear temperature \cite{Gilbert+65,KU}. 
The calculation of delayed neutron emission probabilities also have a complex dependence on nuclear masses. 
In QRPA based calculations, such as those of Ref.~\cite{Moller+03}, nuclear masses set the particle emission thresholds. 
Assuming the ground state structure does not change upon a variation in mass, the change to neutron emission probabilities comes from either a change in the $\beta$-decay Q-value of the parent nucleus, $Q_{\beta}$, or from a change to the neutron separation energies, $S_n(Z,A)$, of the daughter nuclei. 
In more recent calculations that combine QRPA and HF, so called QRPA-HF, nuclear masses impact $\gamma$-ray competition at each neutron emission stage via the combination of possibilities mentioned above \cite{Kawano+08,Mumpower+15b}. 

The role of nuclear masses in the $r$ process is thus three fold: (1) masses and/or mass differences enter into the 
calculation of all other quantities as mentioned above, (2) mass differences play a fundamental role in determining the 
reaction flow in the case of \nggn \ equilibrium, and (3) mass differences impact astrophysical energy generation. Points 
(1) and (2) are addressed in the sensitivity studies described here.

Full chart mass sensitivity studies have been performed which probe the uncertainties in individual nuclear masses 
and their impact on final abundances \cite{Mumpower+15a,Mumpower+15b}. In these studies, changes to individual nuclear 
masses are propagated consistently by recalculating all relevant quantities. When a nuclear mass of a nucleus ($Z$,$N$) is 
varied it impacts the neutron capture rates of ($Z$,$N$) and ($Z$,$N-1$), the separation energies of ($Z$,$N$) and 
($Z$,$N+1$), the $\beta$-decay rates of ($Z$,$N$) and ($Z-1$,$N+1$), and $\beta$-delayed neutron emission probabilities of 
($Z$,$N$), ($Z-1$,$N+1$), ($Z-1$,$N+2$), up to ($Z-1$,$N+12$). Thus, the impact of an individual mass variation can be felt 
through its effects on any (or all) of these quantities.

\begin{figure}
 \begin{center}
  \includegraphics[width=\textwidth]{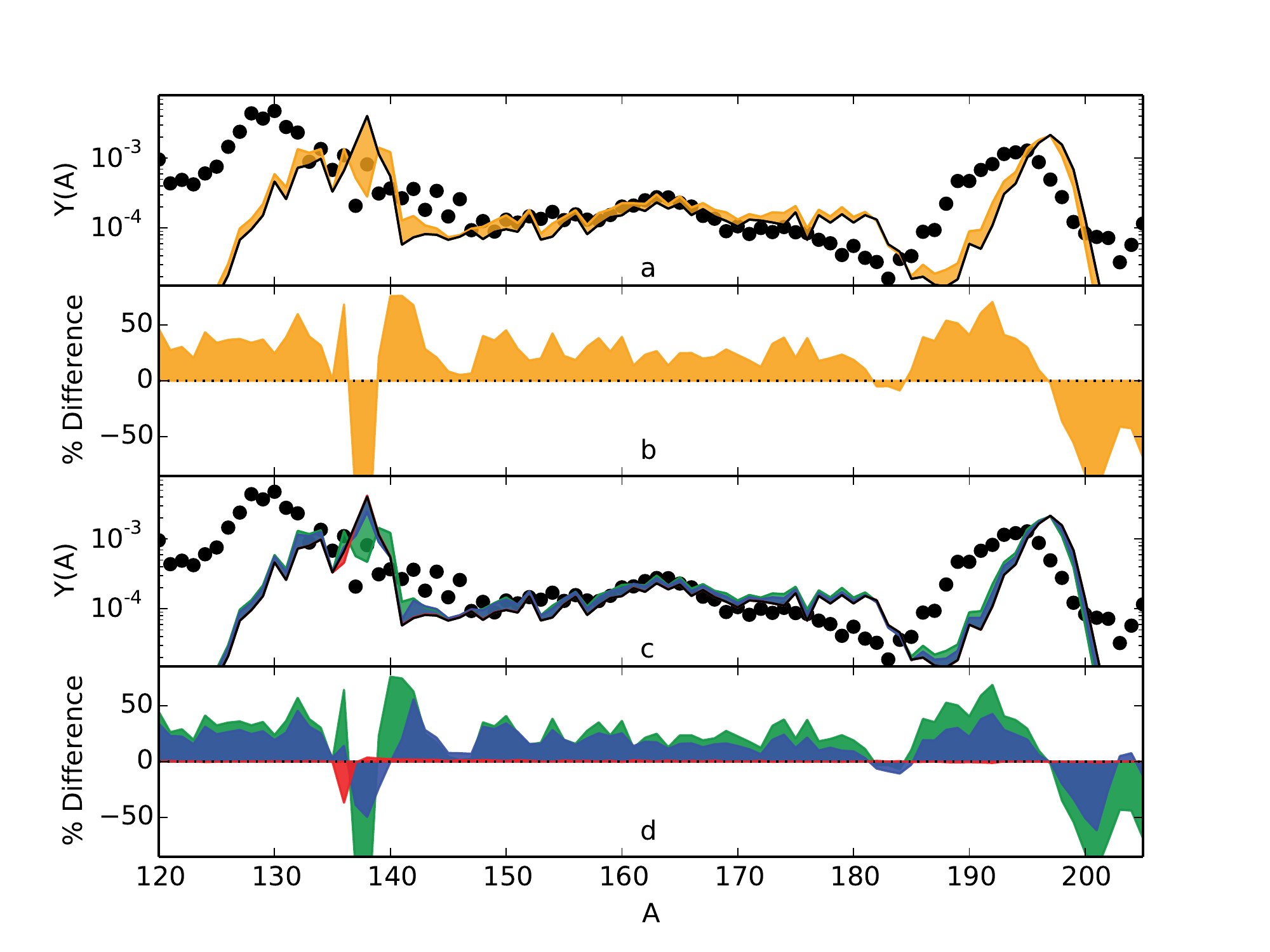}
  \caption{\label{fig:mss-ab} (a) The change in final $r$-process abundances when the mass of $^{140}$Sn is increased by $500$ keV and all dependencies have been considered compared to baseline simulation (black) and solar data (black circles) from \cite{Arlandini+99}. (b) Percent difference of the abundances in (a) to baseline. (c) Separate simulations show the contribution of the mass variation when only propagating changes to neutron captures (red), $\beta$-decays (blue) and photodissociation rates (green). (d) Percent difference of the abundances in (c) to baseline. Simulation data from \cite{Mumpower+15a}. }
 \end{center}
\end{figure}

An example abundance pattern produced when a variation in a single nuclear mass is propagated to all relevant quantities is 
shown by the yellow curve in panels (a) and (b) of Fig.~\ref{fig:mss-ab} for a hot $r$ process. In this simulation the mass 
of $^{140}$Sn is increased by $500$ keV relative to the FRDM1995 calculation, resulting in a large global shift in the final 
abundances compared to the baseline simulation shown in black. Subsequent simulations are shown in panels (c) and (d) that 
separate out the influences of the mass variation on the neutron separation energies, neutron capture rates, and 
$\beta$-decay properties. 

In this case, the propagation of the mass change to photodissociation rates via the neutron separation energies has the 
largest impact on the final abundances (green). $^{140}$Sn is the most populated tin isotope during \nggn \ equilibrium in 
this hot $r$-process example, and so a change to its neutron separation energy acts to shift the $r$-process path slightly, 
to $^{142}$Sn. $^{142}$Sn has a shorter halflife than $^{140}$Sn, so the separation energy change results in increased 
nuclear flow out of the highly abundant tin isotopic chain, leading to changes throughout the pattern. This effect is 
described carefully in early mass sensitivity studies \cite{Brett+12,Aprahamian+14} that considered mass variations 
propagated only to the photodissociation rates.

The early studies \cite{Brett+12,Aprahamian+14} missed the impact of the masses on the weak decay properties and neutron 
capture rates, shown in blue and red, respectively, in the bottom panel of Fig.~\ref{fig:mss-ab} for this example case. The 
mass variation leads to an approximately 45\% increase to the halflife of $^{140}$Sn and a change to the neutron capture 
rate by just under a factor of 2. This halflife change also acts to increase the reaction flow through the tin isotopes in 
the equilibrium phase, pushing material to higher $A$. The impact of the neutron capture rate change is significantly 
smaller; in this hot $r$ process, capture rates are not important until the freezeout phase, by which time the very 
neutron-rich nuclei, including $^{140}$Sn, along the equilibrium $r$-process path have been depopulated as the path shifts 
toward stability.

Figure \ref{fig:mss-ab} clearly illustrates that uncertainties in even a single nuclear mass, on the order of the rms value of most mass models, can greatly influence final $r$-process abundances. 
This result reinforces the conclusions of global Monte Carlo studies which show that rms uncertainties must be reduced in order to predict the finer details of $r$-process abundances. 

\begin{figure}
 \begin{center}
  \includegraphics[width=160mm]{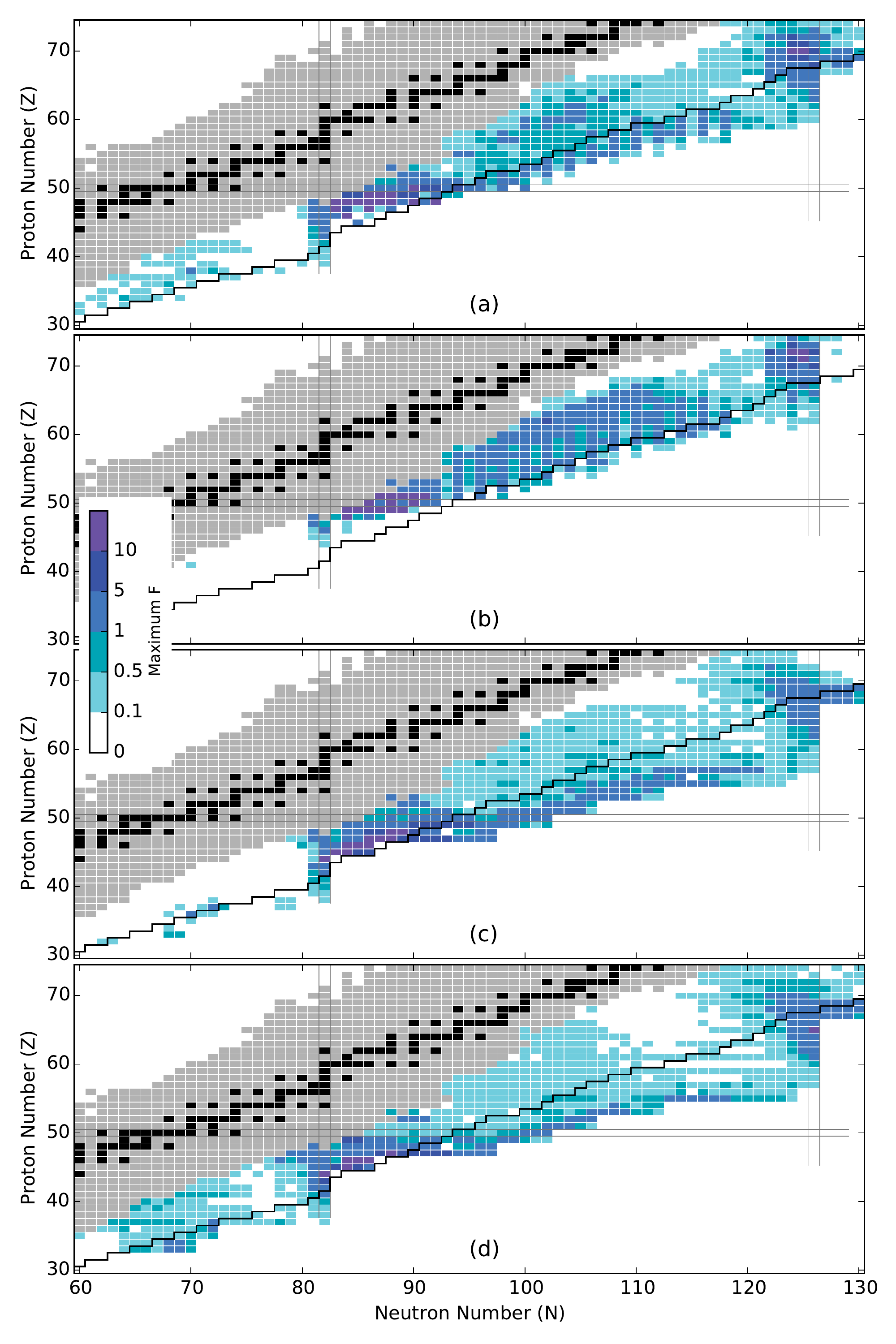}
  \caption{\label{fig:mass-fgrid} Nuclei that significantly impact final $r$-process abundances in four astrophysical conditions (a) low 
entropy hot wind, (b) high entropy hot wind, (c) cold wind and (d) neutron star merger. More influential nuclei are shaded darker, denoting a 
larger maximum impact parameter, $F$ from the $\pm500$ keV mass uncertainty. Light gray shading denotes the extent of measured masses from 
the latest AME with stable nuclei colored black. Estimated neutron-rich accessibility limit shown by a black line for FRIB with intensity of 
$10^{-4}$ particles per second \cite{FRIB}. Simulation data from \cite{Mumpower+15b}. }
 \end{center}
\end{figure}

The final results of four mass sensitivity studies \cite{Mumpower+15b} with consistently calculated nuclear properties are shown in Fig.~\ref{fig:mass-fgrid}. 
The astrophysical conditions chosen are the low entropy hot wind, high entropy hot wind, cold wind, and neutron star merger trajectories described in Sec.~\ref{sec:astro}. 
The intensity of the shading of each nucleus represents the largest $F$ value obtained between the $\pm500$ keV variation from the calculated FRDM2012 mass for each study.

In all scenarios, a number of the nuclei with the most influential nuclear masses lie along the early-time $r$-process path. 
For the hot and neutron star merger scenarios, the $r$ process is in \nggn \ equilibrium during this time, and the influence of the masses is as described above for the $^{140}$Sn example. 
For the cold $r$ process, photodissociation quickly becomes unimportant, and so most of the effect of the masses is through the neutron capture rates and decay properties. 
The bulk of the pattern is built up while neutron capture competes with $\beta$ decay, as shown in Figs.~\ref{fig:timescales} and \ref{fig:ab-evo}; these properties of the nuclei most populated during this phase thus have the greatest leverage on the final pattern.

Also in all scenarios, the influence of nuclear masses is not limited to the early or equilibrium phase of the $r$ process. 
Nuclei with high $F$ measures are found throughout the region between the early $r$-process path and where masses are known 
experimentally, particularly along the decay paths of closed shell nuclei. These are nuclei populated at late times in the 
$r$ process, in the freezeout phase during which the abundance pattern is finalized (Fig.~\ref{fig:ab-evo}) and key features 
such as the rare earth peak form \cite{Surman+97,Mumpower+12a}. Though the dynamics are dominated at that time by the decay 
to stability, neutron capture continues to play a role as long as neutrons are available. Fig.~\ref{fig:timescales} 
indicates that the $\beta$-delayed neutrons emitted at late times are almost all promptly captured; fission provides an 
additional source of late-time neutrons in the merger case. During freezeout individual neutron capture rates determine 
which nuclei will capture these last remaining neutrons, while individual $\beta$-decay rates and $\beta$-delayed neutron 
emission probabilities govern the details of the nuclear flow back to stability. Nuclei with large $F$ measures in 
Fig.~\ref{fig:mass-fgrid} for this region of the nuclear chart have masses that impact $r$-process freezeout through their 
leverage on one or more of these properties.

\subsection{$\beta$-decay properties and neutron capture rates}

The impact of individual $\beta$-decay properties and neutron capture rates on the $r$ process is addressed in part in the 
mass sensitivity studies described above. However the uncertainties in these quantities are not just due to the nuclear mass 
inputs. Mass variations of $\pm500$ keV produce changes to $\beta$-decay and neutron capture rates of factors of 2-5 
\cite{Mumpower+15b}. Much larger uncertainties come from unknown nuclear structure information, as discussed in 
Sec.~\ref{sec:nuc}. 

\begin{figure}
 \begin{center}
  \includegraphics[width=160mm]{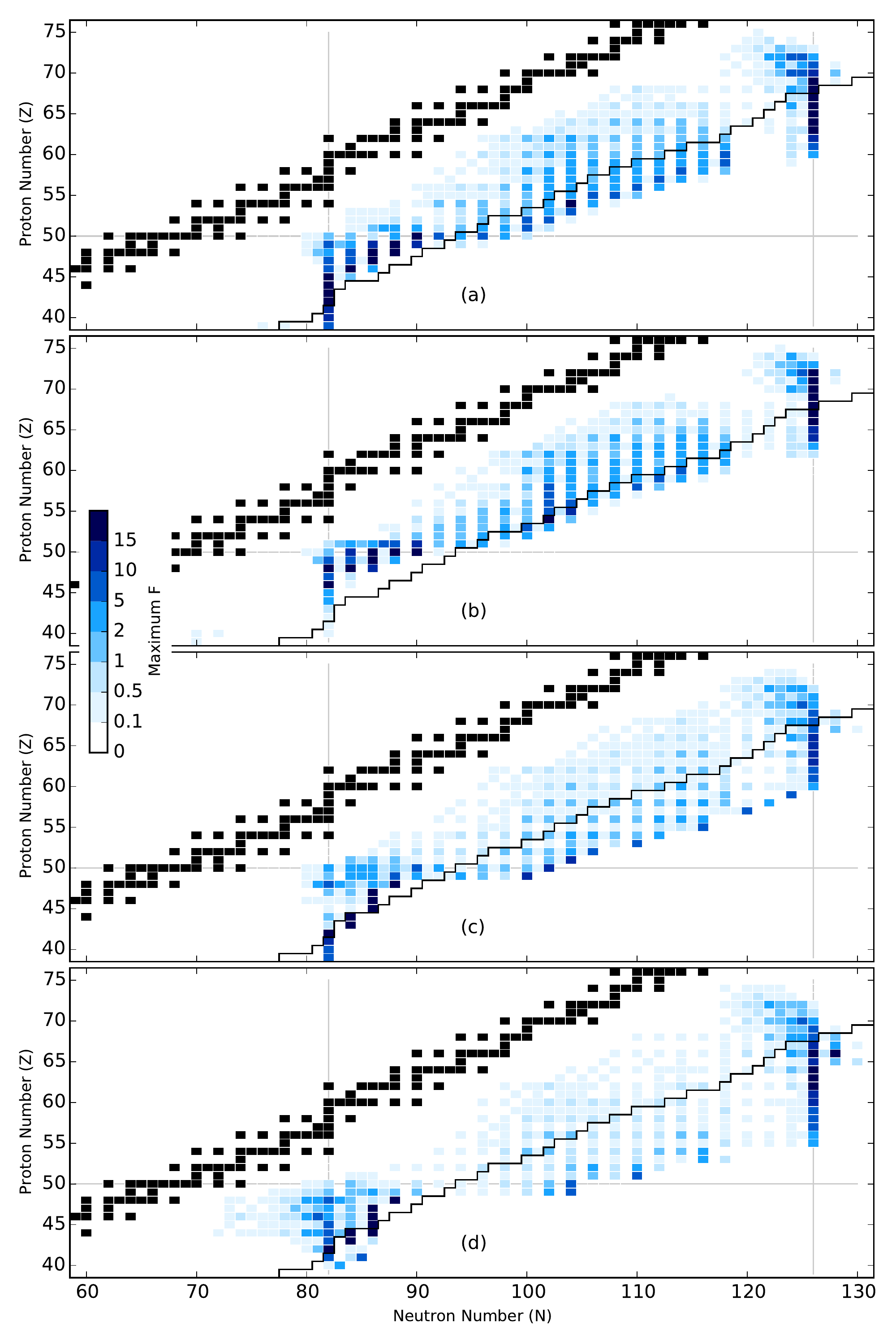}
  \caption{\label{fig:beta-fgrid} Important $\beta$-decay half-lives in four astrophysical environments (a) low entropy hot wind, (b) high entropy hot wind, (c) cold wind and (d) neutron star merger with stable isotopes in black. Estimated neutron-rich accessibility limit shown by a black line for FRIB with intensity of $10^{-4}$ particles per second \cite{FRIB}. Simulation data from \cite{Mumpower+14}.}
 \end{center}
\end{figure}

\begin{figure}
 \begin{center}
  \includegraphics[width=160mm]{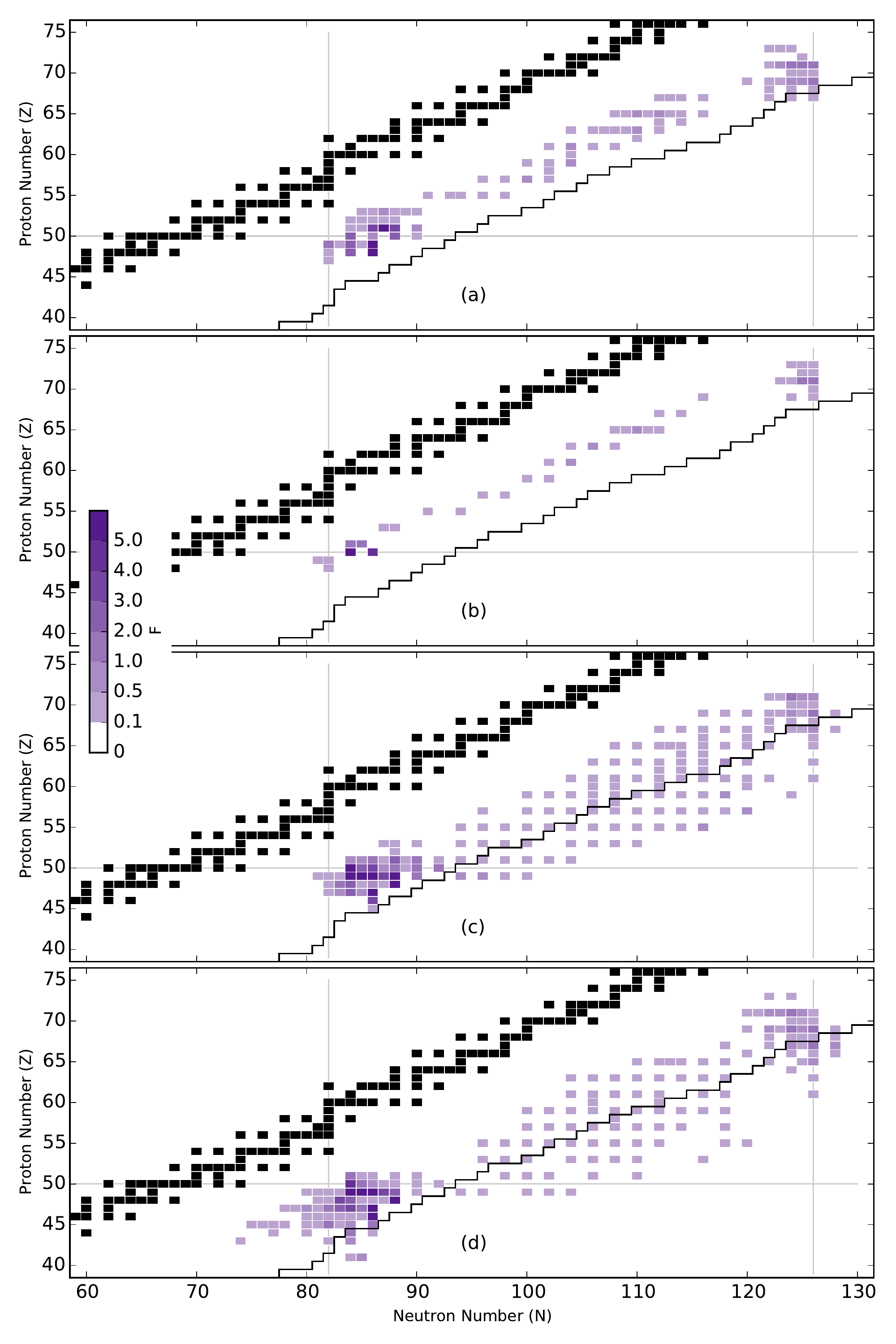}
  \caption{\label{fig:bdne-fgrid} Important $\beta$-delayed neutron emitters in four astrophysical environments (a) low entropy hot wind, (b) high entropy hot wind, (c) cold wind and (d) neutron star merger with stable isotopes in black. Estimated neutron-rich accessibility limit shown by a black line for FRIB with intensity of $10^{-4}$ particles per second \cite{FRIB}. Simulation data from \cite{Surman+15}.}
 \end{center}
\end{figure}

\begin{figure}
 \begin{center}
  \includegraphics[width=160mm]{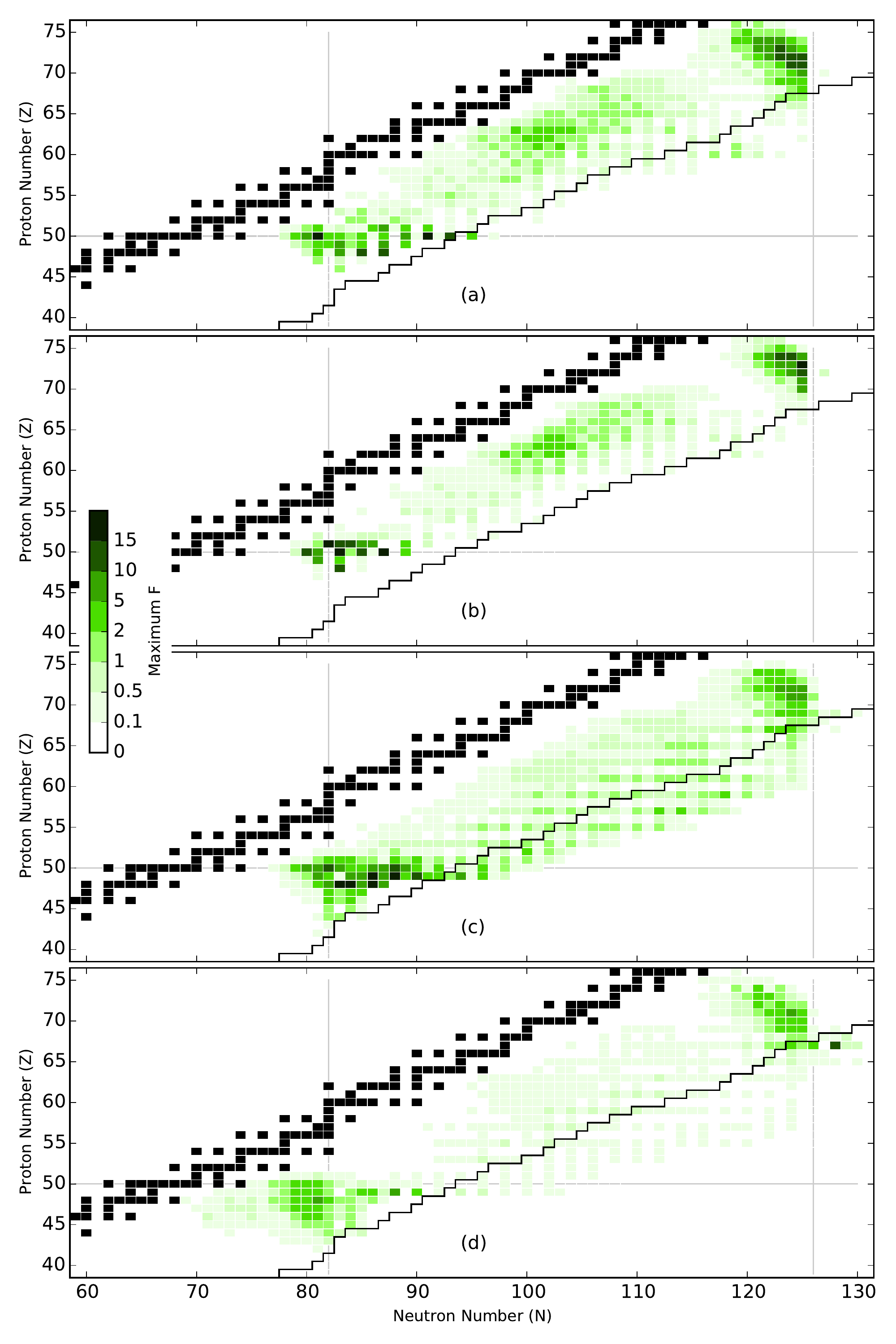}
  \caption{\label{fig:ncap-fgrid} Important neutron capture rates in four astrophysical environments (a) low entropy hot wind, (b) high entropy hot wind, (c) cold wind and (d) neutron star merger with stable isotopes in black. Estimated neutron-rich accessibility limit shown by a black line for FRIB with intensity of $10^{-4}$ particles per second \cite{FRIB}. Simulation data from \cite{Surman+14c}.}
 \end{center}
\end{figure}

Separate sensitivity studies that address these larger uncertainties have been performed for $\beta$-decay rates 
\cite{Surman+14c,Mumpower+14}, $\beta$-delayed neutron emission probabilities \cite{Surman+15}, and neutron capture rates 
\cite{Beun+09,Surman+09,Mumpower+12c,Surman+14c}. Sample results from these studies are shown for $\beta$-decay rates in 
Fig.~\ref{fig:beta-fgrid}, $\beta$-delayed neutron emission probabilities in Fig.~\ref{fig:bdne-fgrid}, and neutron capture 
rates in Fig.~\ref{fig:ncap-fgrid}. They start from the same four sets of astrophysical conditions as in 
Fig.~\ref{fig:mass-fgrid}, but use slightly different nuclear inputs, as the studies all predate FRDM2012. The masses used 
are from FRDM1995, $\beta$-decay properties from \cite{Moller+03}, and neutron capture rates from \cite{NONSMOKER}.

\begin{figure}
 \begin{center}
  \includegraphics[width=\textwidth]{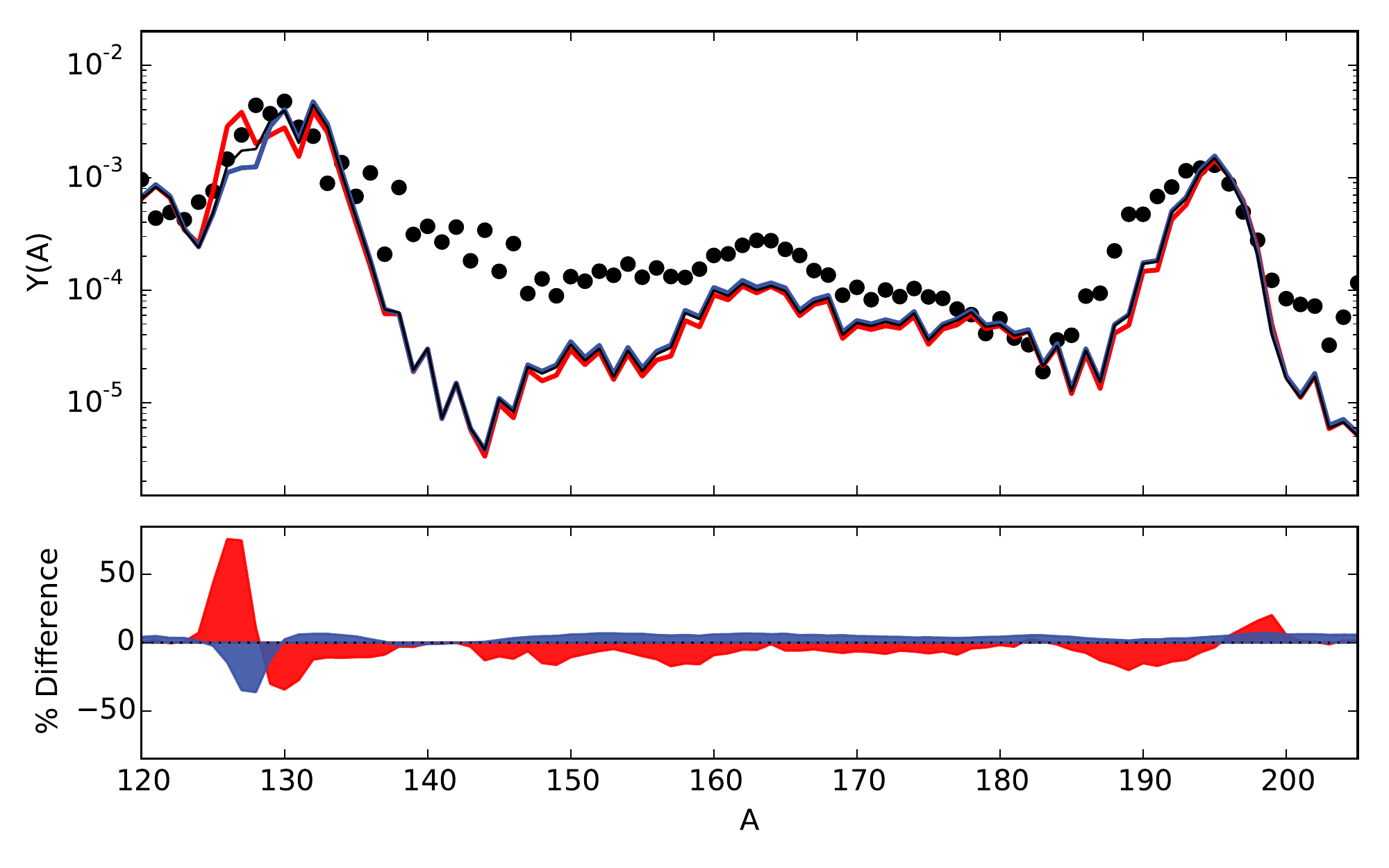}
  \caption{\label{fig:ab-beta-f} The influence of an uncertain $\beta$-decay rate for $^{128}$Ru on the final abundance pattern in a neutron star merger $r$ process. Impact parameters are $F=20.5$ for decreasing the rate (red) and $F=7.56$ for an increase rate (blue). Both the decrease and increase in the rate were by a factor of 10 respectively.}
 \end{center}
\end{figure}

The $\beta$-decay rate studies from \cite{Surman+14c,Mumpower+14} and shown in Fig.~\ref{fig:beta-fgrid} consider factors of 
10 variation in individual $\beta$-decay rates. In all astrophysical scenarios, the most important 
$\beta$-decay rates are for nuclei along the early-time $r$-process path. This is expected since these rates set the 
relative abundances of nuclei along the path via the steady $\beta$ flow condition, 
$\lambda_{\beta}(Z,A_{path})Y(Z,A_{path})\sim$ constant, an approximation that holds remarkably well in all four 
astrophysical scenarios \cite{Mumpower+14}. An example of how an individual $\beta$-decay rate of an $r$-process path 
nucleus influences the $r$-process pattern is shown in Fig.~\ref{fig:ab-beta-f}. $\beta$-decay rates remain important during 
the freezeout phase for as long as neutron capture continues to operate. As a result, modest $\beta$-decay sensitivities are 
seen in Fig.~\ref{fig:beta-fgrid} for nuclei throughout the region in between the path and stability, with somewhat larger 
values for nuclei along the decay paths from the closed shells and in the rare earth region.

Dynamical $r$-process calculations show that $\beta$-delayed neutron emission plays two critical roles during freeze-out: it 
provides an additional source of free neutrons that the rest of the system can use for capture after neutron exhaustion, and 
along with late-time neutron capture it is important for finalizing the details of isotopic abundances, $Y(A)$. For the 
former, the effect of any one $P_{n}$ value is vanishingly small. Sensitivity studies of individual $\beta$-delayed neutron 
emission probabilities therefore capture primarily the latter effect. The studies published to date \cite{Surman+15} examine 
the impact of setting individual $P_{n}$ values to zero; sample results from this work are shown in 
Fig.~\ref{fig:bdne-fgrid}. Note that the $F$-measures here are on a different scale, and most are quite small. This is 
because single $P_{n}$ values only impact the fine details of the abundance pattern, at the very last stages of the $r$ 
process. Thus the global $F$ measure used in \cite{Surman+15} and Fig.~\ref{fig:bdne-fgrid} is not well suited for this type 
of sensitivity study. Upcoming work will examine alternative local sensitivity measures as well as increases to $P_{n}$ 
values.

\begin{figure}
 \begin{center}
  \includegraphics[width=\textwidth]{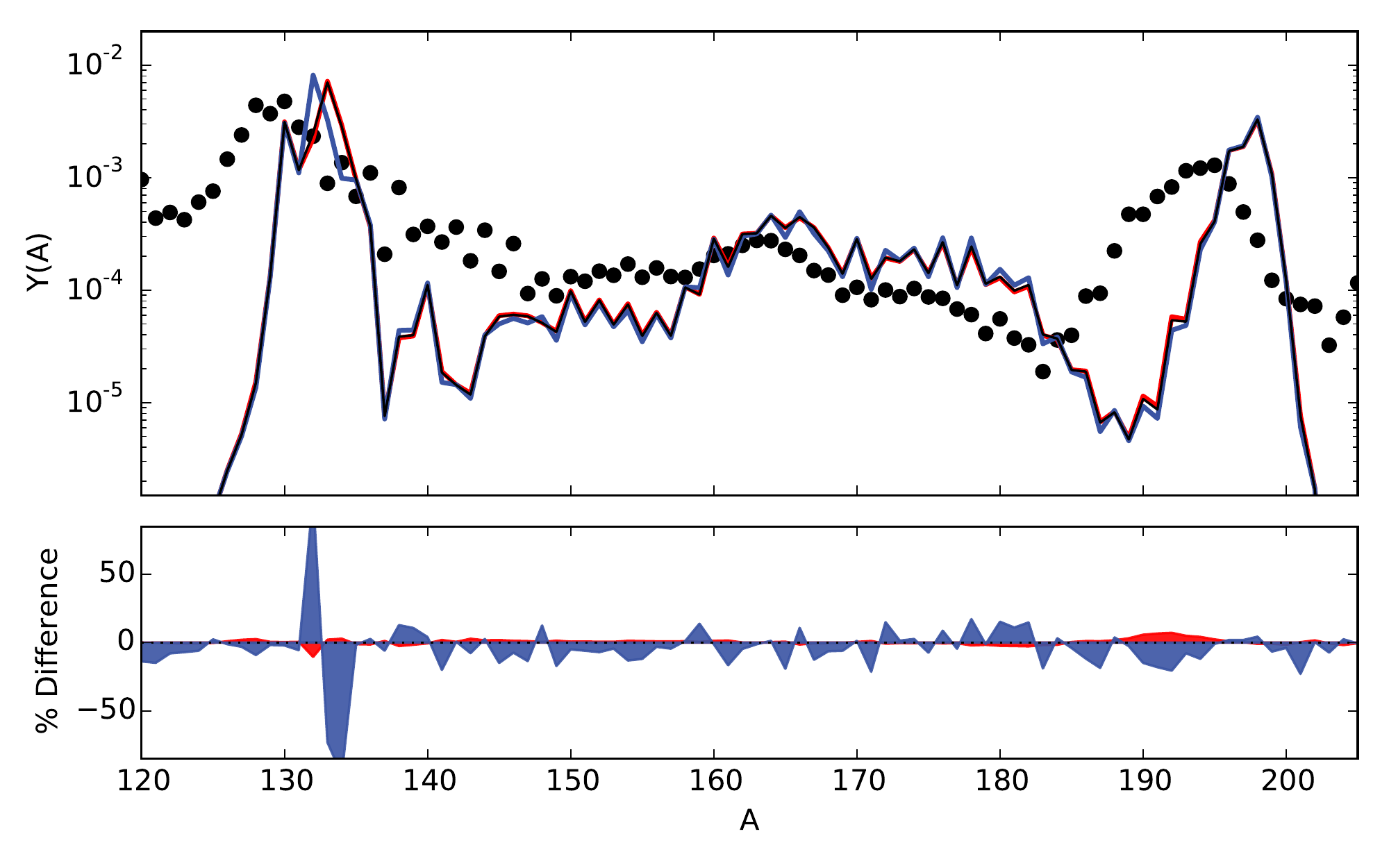}
  \caption{\label{fig:ab-ncap-f} The influence of an uncertain neutron capture rate for $^{133}$Sn on the final abundance pattern in a hot wind $r$ process. Impact parameters are $F=1.77$ for decreasing the rate (red) and $F=20.6$ for an increase rate (blue). Both the decrease and increase in the rate were by a factor of 100 respectively.}
 \end{center}
\end{figure}

Neutron capture rates are arguably the most uncertain of nuclear data inputs for the $r$ process, with few experimental 
constraints and theoretical predictions that differ by orders of magnitude, as illustrated in Fig.~\ref{fig:ncaptheo}. The 
neutron capture rate sensitivity studies of \cite{Beun+09,Surman+09,Mumpower+12c,Surman+14c} and shown in 
Fig.~\ref{fig:ncap-fgrid} examine variations to the rates of factors of 100. These rate changes have no impact 
for simulations in \nggn \ equilibrium, and as a result the sensitivity measures for nuclei along the hot $r$-process path 
are essentially zero, shown in the top two panels of Fig.~\ref{fig:ncap-fgrid}. Individual neutron capture rates along the 
early-time $r$-process path are influential in the cold and merger cases, since in both of these scenarios the neutron 
abundance is still quite high when photodissociation becomes negligible. In all cases, individual neutron capture rates 
shape the details of the abundance pattern throughout the freezeout phase, for as long as there are neutrons available to 
capture. Thus nuclei with modest neutron capture sensitivity measures extend almost to stability in all four scenarios, with 
the greatest $F$ measures concentrated along the decay pathways of closed shell nuclei and in the rare earth region, where 
they impact the formation of the rare earth peak \cite{Mumpower+12c}. An example of the impact of a single neutron capture 
rate variation on the final pattern for a hot wind $r$ process is included in Fig.~\ref{fig:ab-ncap-f}.

The complete sensitivity study results of Figs.~\ref{fig:mass-fgrid}, \ref{fig:beta-fgrid}, \ref{fig:bdne-fgrid}, \& \ref{fig:ncap-fgrid} are included in a table in the appendix. 
Sensitivity measures $F$ are stated explicitly wherever $F>0.01$; an asterisk is used to indicate $0<F<0.01$. 
Since the studies of masses started from a different baseline nuclear data set than the other studies, the sets of nuclei included in each study do not necessarily overlap. 
If a nucleus is not included in a particular study it is indicated by a dash in the table. 

Some caveats regarding the table:

\noindent
(1) All studies are for main $A>120$ $r$ processes. $F$ measures are listed for some nuclei with $A<120$; these indicate the impact of their nuclear properties on $A>120$ nucleosynthesis only. 

\noindent
(2) The sensitivity measures $F$ are {\bf global} measures. The largest measures correspond to either large local abundance pattern changes or somewhat smaller changes throughout the pattern, or some combination of the two. 
Modest local changes are not captured by this measure.

\noindent
(3) The four sets of astrophysical conditions chosen for the studies are meant to be representative of a variety of possible scenarios but are by no means exhaustive. 
Different astrophysical conditions will produce different $F$ measures, and some nuclei labeled with asterisks in the table may be important in other scenarios, or with a different choice of baseline nuclear data. 

\section{Radioactive ion beam facilities, new techniques, and the future}\label{sec:exp}

Detangling the astrophysical origins of the elements involves understanding the nuclear properties of thousands of nuclei in regions 
of the chart of nuclides where information is highly limited or unknown \cite{Wiescher+15}. To ameliorate the situation, there has 
been a global competition towards the construction and development of facilities and techniques to measure the properties of nuclei 
far from stability. For the $r$ process, properties such as nuclear masses, $\beta$-decay rates, neutron capture rates, and 
$\beta$-delayed neutron emission probabilities are of key importance, as described above. The advances that have led to the global 
development of the existing radioactive ion beam facilities and future construction of new facilities have and will further transform 
the study of nuclei and nuclear properties \cite{Blumenfeld+13}.

There are presently two techniques for the production of the most neutron-rich nuclei: (1) the ISOL (Isotope Separation On-Line 
method) and (2) the in-flight projectile fragmentation approach. The TRIUMF laboratory in Canada, the ISOLDE experiment at CERN in 
Switzerland, the University of Jyv{\"a}skyl{\"a} center of excellence in Finland, and SPIRAL2 at the GANIL laboratory in France have 
implemented the ISOL method of using protons or light ions on a heavy target such as Uranium to produce radioisotopes, while 
spontaneous fission products from a Californium source are extracted at the CARIBU facility at Argonne National Laboratory in the USA. The 
fragmentation method on the other hand is the choice for the NSCL at Michigan State University in the USA, for GSI and the future GSI-FAIR 
project in Germany, the RIBLL project \cite{Sun+03} in China, the RIBF facility at RIKEN in Japan, and the future FRIB at Michigan 
State University in the USA. In-flight projectile fragmentation involves collisions of nuclei at relativistic energies to produce the 
highly radioactive unstable fragments. The project in KORIA \cite{Moon+14} plans to use both approaches of ISOL and fragmentation. 
There are advantages and disadvantages to both methods with the fragmentation approach allowing access to shorter half-lives while the 
ISOL method depending on the chemistry involved can produce more intense beams of potentially higher quality.

The essential properties of interest to the $r$-process are as listed previously: nuclear masses, $\beta$-decay rates, neutron capture 
rates, and $\beta$-delayed neutron emission probabilities. The beam production rates that are required for the measurements of these 
properties vary over orders of magnitude. For example, it is possible to make measurements of half-lives or masses in traps where a 
good separation is achieved by tens of events. Below, we discuss the current approaches and techniques in use to measure masses of 
nuclei, $\beta$-decay properties, and neutron capture rates.

\subsection{Mass measurements}
Recent advances in techniques to measure nuclear masses with Penning traps \cite{Lunney+03} and ion storage rings \cite{Lunney+03, 
Blaum+06} have resulted in a significant increase in the numbers of nuclei whose masses have been measured. Penning traps in 
particular are responsible for the significantly large percentage of the increase in the numbers of measured masses for the AME2012 
evaluation with high precision. The advent of radioactive ion beam facilities will allow measurements of many more nuclear masses, 
including those of interest for the $r$ process. Figure \ref{fig:exp-reach} shows the sum of the mass sensitivity studies in four 
astrophysical scenarios from Fig.~\ref{fig:mass-fgrid}, along with information on the experimental reach of current and future 
facilities. The color intensity shown in this plot indicates the importance of a given nucleus in any one of the four astrophysical 
scenarios. The solid line is the calculated production limit with FRIB. Superimposed on this figure are some measurements made at two 
ISOL facilities: TRIUMF (ISAC) and CARIBU at Argonne National Laboratory. At ISOL facilities, the production rates that allow 
measurements depend on a number of complicated issues varying from source strengths, targets, and various other extraction conditions. 
We have therefore chosen to illustrate a set of nuclei that have been produced with some nominal yields as measured at TRIUMF 
(circles) and a number of mass measurements made at CARIBU (stars) \cite{VanSchelt+13}. In addition, we show the recent measurements 
of $\beta$-decay rates for $110$ nuclei \cite{Lorusso+15} from RIBF at RIKEN in Japan using the WAS3ABi array of eight double sided 
silicon strip detectors in coincidence with the 84 high purity germanium EURICA array \cite{Soderstrom+13}. These results from three 
facilities around the world indicate the present reach of radioactive ion beam experiments in a schematic way while the FRIB solid 
black line indicates the potential future experimental reach. New techniques are bound to be developed to allow access to measurements 
of neutron rich radioactive ion beam nuclei with even lower yields than estimated here.

\subsection{$\beta$-decay properties}
\begin{figure}
 \begin{center}
  \includegraphics[width=\textwidth]{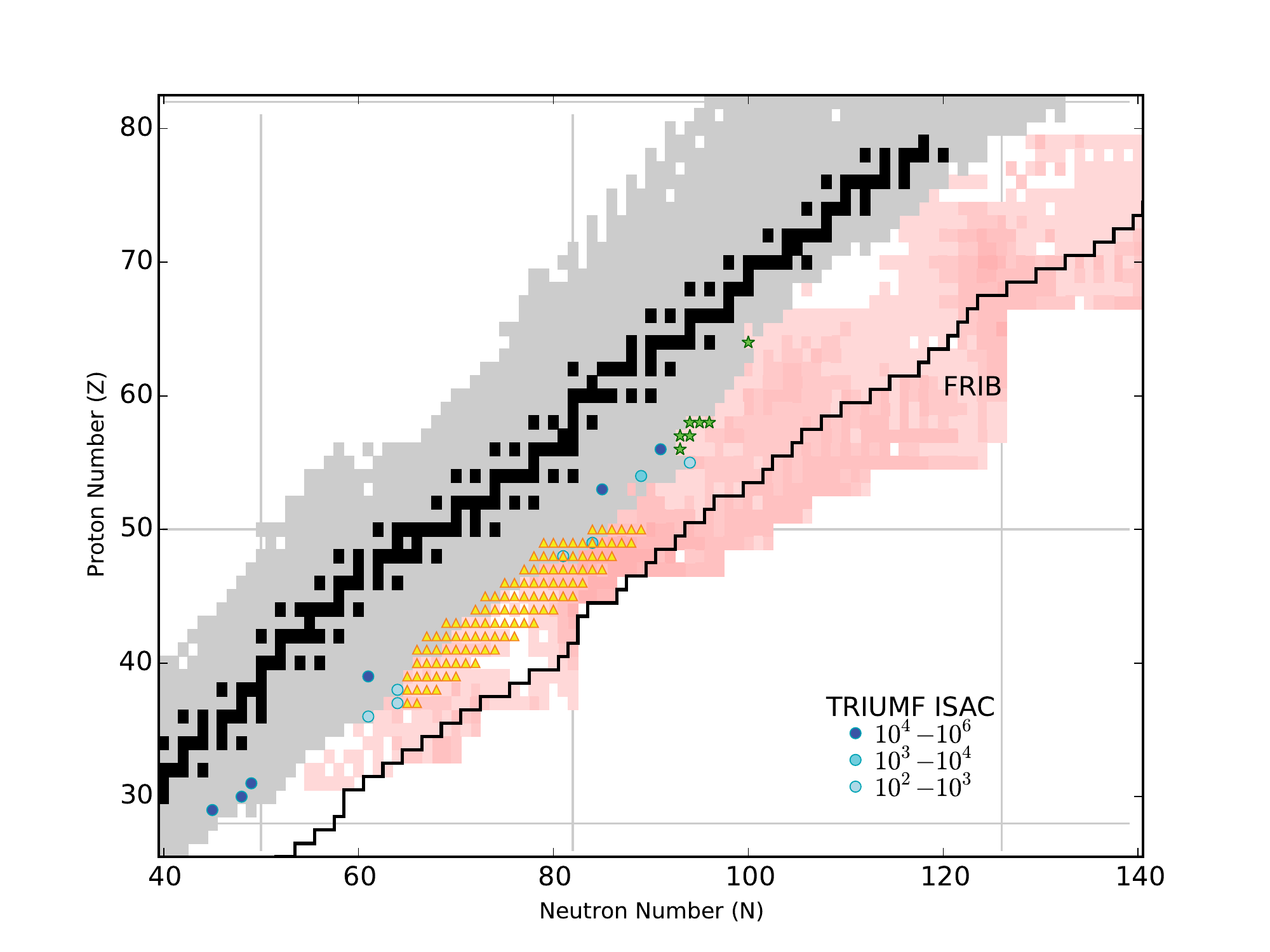}
  \caption{\label{fig:exp-reach} Recent measurements on neutron-rich nuclei and estimated capabilities at the future FRIB (black line: $10^{-4}$ particles per second \cite{FRIB}). Triangles denote recently measured $\beta$-decay half-lives @ RIKEN \cite{Lorusso+15}, stars represent recently measured masses with the CPT at CARIBU \cite{VanSchelt+13}, and circles are the nominal production rates measured for these nuclei with ISAC at TRIUMF \cite{TRIUMF_ISAC}. Background shading denotes the extent of the latest AME2012 evaluation of masses (light gray), influential $r$-process nuclei from \cite{Mumpower+15b} with $F\geq0.1$ (pink), and stable nuclei (black). }
 \end{center}
\end{figure}

The experimental reach suggested in Fig.~\ref{fig:exp-reach} is as relevant for $\beta$ decay as for masses. In addition to new 
halflife measurements, $\beta$-delayed neutron emission probabilities will be the focus of a number of experimental campaigns. A 
recent IAEA report \cite{Abriola+11,Dillmann+11} states that there are approximately $200$ $\beta$-delayed neutron emitters that have 
been measured over the entire chart of nuclides where over $75$ of them are below $A<70$. Measurements of these probabilities are 
particularly challenging due to the experimental complications of detecting neutrons and $\beta$s in high efficiency and good resolution. 
There are many ongoing developments to augment the experimental measurements. Some current examples include the BELEN detector being 
developed (Beta deLayEd Neutron detector) for use at GSI/FAIR and already implemented for measurements at Jyv{\"a}skyl{\"a} 
\cite{Gomez-Hornillos+11}. New approaches avoid the detection of the neutron itself by measuring the recoil of the neutron 
\cite{Yee+13} where a Paul trap is used to measure the ion recoils in coincidence with the $\beta^-$ particles. There are still others 
such as the Hybrid 3HEN \cite{Grzywacz+14}, MONSTER (Modular Neutron time-of-flight SpectromeTER \cite{Martinez+14,Tain+15}, NERO 
(Neutron Emission Ratio Observer) \cite{Pereira+09} in coincidence with the BCS (beta counting station) at the NSCL, and SIMBA 
(Silicon Implantation and Beta Absorber) \cite{Caballero+13} used at the GSI fragment separator facility to measure $\beta$-decay and 
$\beta$-delayed neutron emission probabilities beyond $N=126$.
 
\subsection{Neutron capture rates} 
Measurements of neutron capture rates on radioactive nuclei far from stability pose exceptionally difficult challenges in addition to 
those of mass and decay property measurements.  The difficulties here lie in the need for significantly more intense beams on 
radioactive targets which are limited in production and very short lived. This aspect forbids the direct measurements of capture rates 
and leads to the use of completely theoretical estimates for the capture rates in $r$-process simulations. There are however a number 
of promising \emph{indirect} approaches to measuring neutron capture cross sections for nuclei that are under development. One method 
is the surrogate reaction technique developed in the 1970's by Ref.~\cite{Cramer+70}. This very general technique involves determining 
cross sections for nuclear reactions that proceed through a compound nucleus via a surrogate reaction which also populates the same 
compound nucleus but is much easier to measure \cite{Cizewski+07,Bardayan+13,Escher+12}. Since the compound nucleus does not have 
`memory' of the formation process, details of the desired reaction channel may be extracted using a number of methods including 
inelastic scattering, neutron transfer or pick-up reactions. In the case of neutron capture, $(d,p)$ reactions have been used as a 
surrogate \cite{Jones+11,Kozub+12}. A second promising approach to measuring neutron capture cross sections involves the so called 
$\beta$-Oslo method \cite{Spyrou+14}. In this approach high-lying levels in the nucleus of interest are populated via $\beta$ decay. A 
total absorption spectrometer is used to measure $\gamma$-rays and thereby determine the level density as well as the $\gamma$-ray 
strength function experimentally. These measurements can then be combined with theoretical calculations of an optical model 
potential to derive the neutron capture cross section. The power of this method is that the same experimental technique can also be 
used for measurements of $\beta$-decay properties.

\subsection{Future outlook}

Future measurements have the promise to resolve much of the uncertainties that go into $r$-process simulations. With nuclear physics 
uncertainties reduced, e.g. Fig.~\ref{fig:ab-err_mass}, details of the abundance pattern can be clearly resolved and thus used to 
constrain astrophysical conditions. One example of an abundance pattern feature that can point to the site of the $r$ process is the 
rare earth peak ($A\sim160$).

The rare earth peak found in the solar $r$-process residuals occurs away from closed neutron shells and thus must form by 
a unique mechanism. Two possibilities for peak formation are either dynamically during freezeout \cite{Surman+97,Mumpower+12a} or via 
fission recycling \cite{Steinberg+78,Goriely+13}. In the dynamical mechanism, peak formation is thought to occur when the $r$-process 
path encounters a deformation maximum (or other region of enhanced stability) during the decay back to stability. The location and 
size of the peak produced in this way is sensitive to both the nuclear properties of nuclei away from stability in the rare earth 
region and the temperature and density evolution of the environment \cite{Mumpower+12b}. In the fission recycling mechanism, the peak 
is thought to form from fission products deposited in the $A\sim 160$ region. This method of formation requires very neutron-rich 
conditions, and the placement of the peak depends on where the $r$ process terminates via fission as well as the fission product 
distributions of the nuclei near the termination point.

\begin{figure}
 \begin{center}
  \includegraphics[width=\textwidth]{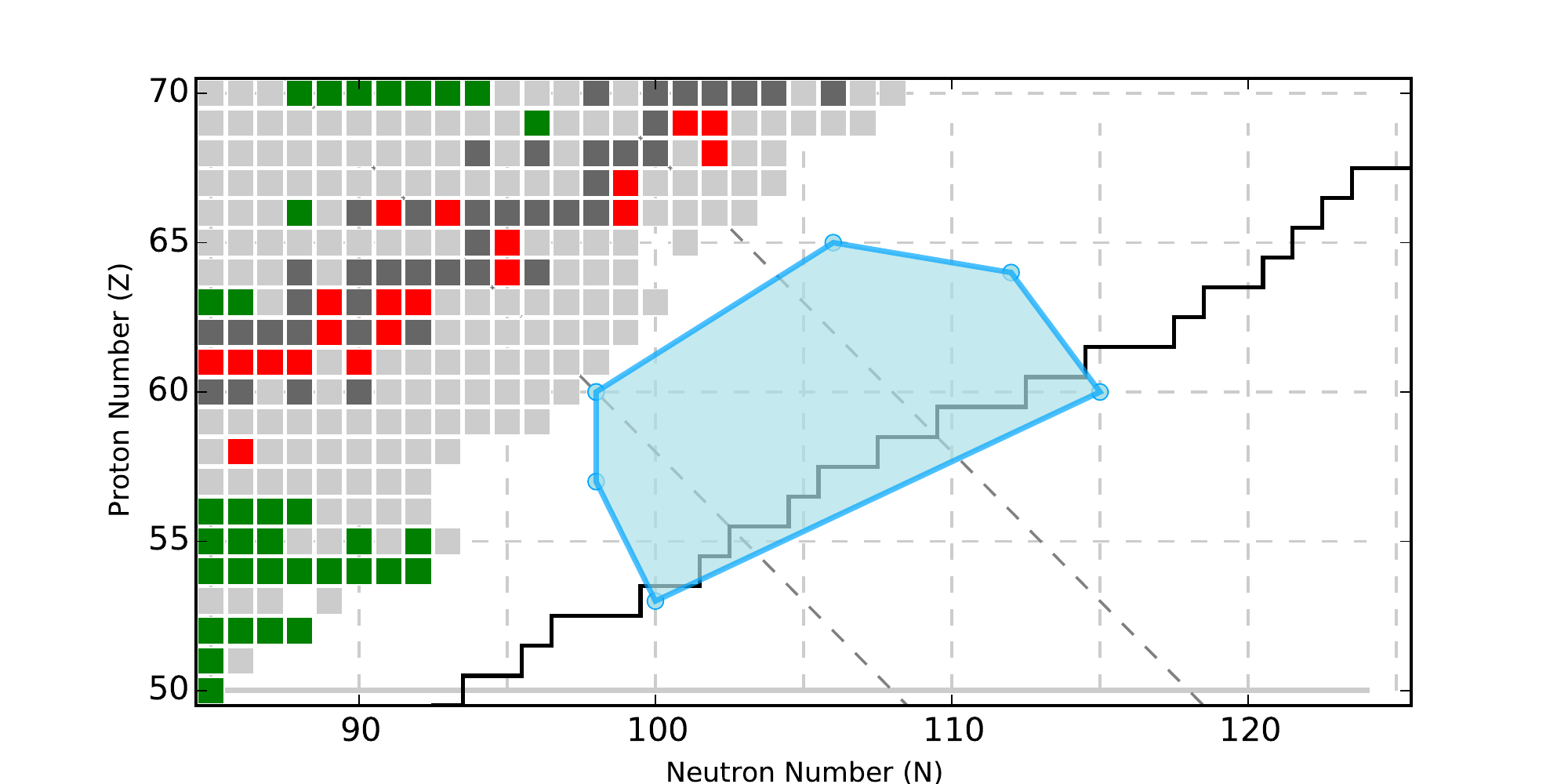}
  \caption{\label{fig:ree} Nuclei in the region of interest for the formation of the rare earth peak (blue). Mass measurements (AME2003) shown in gray with more recent trap measurements highlighted in green. Red denotes extent of known neutron capture cross sections in the region while dark gray squares denote stable nuclei. Figure adapted from Ref.~\cite{Mumpower+12a}, with the FRIB limit from Fig.~\ref{fig:exp-reach} added in black. }
 \end{center}
\end{figure}

The nuclear physics of the fission fragment explanation is not directly testable, as the relevant properties are fission barrier 
heights and product distributions for nuclei close to the neutron drip line above the $N=126$ closed shell, well out of 
experimental reach. However, future facilities, such as FRIB, will be able to access most of the masses and half-lives in the 
rare earth region, as depicted in Fig.~\ref{fig:ree}. These measurements are crucial in elucidating the dynamic formation 
mechanism of the rare earth peak. Signatures such as the `kink' in one neutron separation energies or enhanced stability in 
half-lives will indicate the existence of the proposed deformation maximum in the region. If a region of local enhanced 
stability is found then the peak formation process itself can be used to constrain astrophysical conditions for a main $r$ 
process \cite{Mumpower+12b}. If a region of local enhanced stability is not found then this indicates the peak should be formed 
by fission recycling. This would imply the existence of very specific fission fragment distributions, from which we can learn 
about the nuclear structure of the fissioning nuclei. It would also suggest that a main $r$ process must achieve extremely 
neutron-rich conditions, which would point very strongly to a neutron star merger environment as the primary site of the $r$ 
process.

\subsection{Summary}

Many of the thousands of nuclei important for the $r$ process will be accessible in upcoming facilities. Ideally, given enough 
time and enough resources, one should reach for measurements of all of them. This review attempts to summarize the impact these 
nuclear physics inputs have on the predictions of $r$-process abundances and points to the most influential nuclei to be 
measured at present and future radioactive ion beam facilities worldwide with the potential to lead the way towards answering 
one of the grand challenges in all of physics.

We caution that this review, however, does not provide an exhaustive list. The set of astrophysical trajectories considered 
cover the range of the most popular potential $r$-process sites, though the $r$-process conditions that nature has chosen could 
be quite different from these. Individual fission properties and neutrino interaction rates may also be of key importance but 
have not yet been investigated with the same type of sensitivity analysis. Finally here we have focused primarily on the 
properties of specific nuclei that made the largest magnitude change to the overall abundance pattern. Local changes can be 
equally important and can weigh more heavily nuclei that are closer to stability, as in the rare earth region as described 
above. Additionally, nuclei with small and negligible direct impact on the $r$ process as shown by our $F$-metric can still 
point to nuclear structure trends far away from stability that can be important for building realistic nuclear models. It is 
only with coordinated efforts in nuclear experiment and theory, along with advances in astrophysical modeling and observations, 
that a solution to the longstanding mystery of the $r$-process astrophysical site will be found.

\section{Acknowledgements}

We thank Peter M{\"o}ller, Dong-Liang Fang, Scott Marley, Mary Beard, Andrew Steiner, Ian Bentley \& Maxime Brodeur for helpful discussions. 
This work was supported in part by the National Science Foundation through the Joint Institute for Nuclear Astrophysics grant numbers PHY0822648 and PHY1419765, and the Department of Energy under contracts DE-SC0013039 (RS), DE-FG02-02ER41216 (GCM) \& DE-FG02-10ER41677 (GCM).

\newpage
\section{Appendix}
\begin{longtable}{@{\extracolsep{\fill}} ccc | cccc | cccc | cccc }
\caption{\label{tab:F} Important nuclei from sensitivity studies given criteron $F_{max}\geq0.1$ for at least one astrophysical condition: (a) low entropy hot wind, (b) high entropy hot wind, (c) cold wind and (d) neutron star merger. Asterisk denotes $F_{max}$ below threshold of $0.01$. Dash denotes nucleus was not in the study. }
\setlength{\tabcolsep}{2cm}\\ 
\multicolumn{3}{c}{Nucleus} & \multicolumn{4}{c}{$\beta$-decay} & \multicolumn{4}{c}{Neutron capture} & \multicolumn{4}{c}{Nuclear mass}\\ 
\hline\hline
$Z$ & $N$ & $A$ & $F^{a}_{max}$ & $F^{b}_{max}$ & $F^{c}_{max}$ & $F^{d}_{max}$ & $F^{a}_{max}$ & $F^{b}_{max}$ & $F^{c}_{max}$ & $F^{d}_{max}$ & $F^{a}_{max}$ & $F^{b}_{max}$ & $F^{c}_{max}$ & $F^{d}_{max}$\\ 
\hline
\endhead
$30$ & $49$ & $79$ & $\ast$ & $0.03$ & $0.08$ & $\ast$ & $\ast$ & $0.08$ & $0.20$ & $\ast$ & --- & --- & --- & --- \\ 
$30$ & $52$ & $82$ & $\ast$ & $0.03$ & $0.03$ & $\ast$ & $\ast$ & $\ast$ & $0.22$ & $\ast$ & $\ast$ & $\ast$ & $\ast$ & $\ast$ \\ 
$30$ & $53$ & $83$ & $\ast$ & $\ast$ & $\ast$ & $\ast$ & $\ast$ & $0.06$ & $0.11$ & $\ast$ & $\ast$ & $\ast$ & $\ast$ & $\ast$ \\ 
$30$ & $54$ & $84$ & $0.46$ & $0.40$ & $0.05$ & $\ast$ & $\ast$ & $\ast$ & $0.80$ & $\ast$ & $\ast$ & $\ast$ & $\ast$ & $\ast$ \\ 
$30$ & $55$ & $85$ & $\ast$ & $\ast$ & $0.02$ & $\ast$ & $\ast$ & $0.29$ & $0.03$ & $\ast$ & $\ast$ & $\ast$ & $\ast$ & $\ast$ \\ 
$30$ & $56$ & $86$ & $4.21$ & $0.70$ & $1.00$ & $\ast$ & $\ast$ & $\ast$ & $0.17$ & $\ast$ & $\ast$ & $\ast$ & $\ast$ & $\ast$ \\ 
$30$ & $57$ & $87$ & $\ast$ & $\ast$ & $\ast$ & $\ast$ & $\ast$ & $\ast$ & $0.13$ & $\ast$ & $\ast$ & $\ast$ & $\ast$ & $\ast$ \\ 
$31$ & $48$ & $79$ & $\ast$ & $\ast$ & $0.01$ & $\ast$ & $\ast$ & $0.05$ & $0.13$ & $\ast$ & --- & --- & --- & --- \\
\end{longtable}

\bibliographystyle{unsrt}
\bibliography{refs}

\end{document}